\documentclass[fleqn,usenatbib]{mnras}

\usepackage{newtxtext,newtxmath}

\usepackage[T1]{fontenc}
\DeclareRobustCommand{\VAN}[3]{#2}
\let\VANthebibliography\thebibliography
\def\thebibliography{\DeclareRobustCommand{\VAN}[3]{##3}\VANthebibliography}

\usepackage{graphicx}	
\usepackage{amsmath}	
\usepackage{physics}
\usepackage{subfigure}
\usepackage{color}
\usepackage{gensymb}
\usepackage{hyperref}

\title[Detecting compact objects in GW astronomy]{Resolving dichotomy in compact objects through continuous gravitational waves observation}

\author[Kalita et al.]{
Surajit Kalita,$^{1}$\thanks{E-mail: surajitk@iisc.ac.in}
Tushar Mondal,$^{1}$\thanks{E-mail: mtushar@iisc.ac.in; Present affiliation: International Centre for Theoretical Sciences, Tata Institute of Fundamental Research, Bangalore 560089, India}
Christopher A. Tout,$^{2}$\thanks{Email: cat@ast.cam.ac.uk}
Tomasz Bulik,$^{3}$\thanks{Email: tb@astrouw.edu.pl}
Banibrata Mukhopadhyay$^{1}$\thanks{E-mail: bm@iisc.ac.in}
\\
$^{1}$Department of Physics, Indian Institute of Science, Bangalore 560012, India\\
$^{2}$Institute of Astronomy, The Observatories, Madingley Road, Cambridge CB3 0HA, UK\\
$^{3}$Astronomical Observatory, University of Warsaw, Al. Ujazdowskie 4, 00478 Warszawa, Poland
}

\date{Accepted XXX. Received YYY; in original form ZZZ}

\pubyear{2021}

\begin{document}
\label{firstpage}
\pagerange{\pageref{firstpage}--\pageref{lastpage}}
\maketitle

\begin{abstract}
More than two dozen soft gamma-ray repeaters (SGRs) and anomalous X-ray pulsars (AXPs) have been detected so far. 
These are isolated compact objects. Many of them are either found to be associated with supernova remnants or 
their surface magnetic fields are directly measured, confirming that they are neutron stars (NSs). However, 
it has been argued that some SGRs and AXPs are highly magnetized white dwarfs (WDs). Meanwhile, the 
existence of super-Chandrasekhar WDs has remained to be a puzzle. However, not even a single such massive 
WD has been observed directly. Moreover, some WD pulsars are detected in electromagnetic surveys and some 
of their masses are still not confirmed. Here we calculate the signal-to-noise ratio for all these 
objects, considering different magnetic field configurations and thereby estimate the required time for 
their detection by various gravitational wave (GW) detectors. For SGRs and AXPs, we show that, if these are 
NSs, they can hardly be detected by any of the GW detectors, while if they are WDs, \textit{Big Bang Observer 
(BBO)}, \textit{DECi-hertz Interferometer Gravitational wave Observatory (DECIGO)} and \textit{Advanced Laser 
Interferometer Antenna (ALIA)} would be able to detect them within a few days to a year of integration, 
depending on the magnetic field strength and its configuration. Similarly, if a super-Chandrasekhar WD has a 
dominant toroidal field, we show that even \textit{Laser Interferometer Space Antenna (LISA)} and 
\textit{TianQin} would be able to detect it within one year of integration. We also discuss how GWs can 
confirm the masses of the WD pulsars.
\end{abstract}




\section{Introduction}

The detection of gravitational waves (GWs) in the LIGO/Virgo detectors from various compact object merging 
events has opened a new branch in astronomy. At present, there is enormous effort going into increasing their 
sensitivity to detect continuous GWs (CGWs) emitted from various systems \citep{2019Univ....5..217S}. LIGO and 
Virgo operate at frequencies of greater than about $1\,$Hz. Hence, in the future, these detectors 
will be able to detect CGWs from millisecond neutron star (NS) pulsars and compact binary inspirals 
\citep{2015CQGra..32a5014M}. However, various soft gamma repeaters (SGRs) and anomalous X-ray pulsars (AXPs) 
have been detected over the last few decades 
\citep{2014ApJS..212....6O}\footnote{\url{http://www.physics.mcgill.ca/~pulsar/magnetar/main.html}}. SGRs are 
detected through bursts in hard X-rays or soft gamma-rays, whereas AXPs are detected in soft X-rays 
\citep{2008A&ARv..15..225M}. These are isolated compact objects, mostly believed to be highly magnetized NSs 
\citep{2001ApJ...559..963G}. However, some researchers argue that they could be magnetized white dwarfs (WDs) 
to highly magnetized WDs \citep[B-WDs,][]{2012PASJ...64...56M,2013A&A...555A.151B,2016JCAP...05..007M}. These 
arguments are primarily based on the explanation of their X-ray luminosities. It is also found that their 
X-ray luminosities are $2$ to~$3$ orders of magnitude higher than their spin-down luminosities 
\citep{2014ApJS..212....6O}. They are usually found to be isolated and hence they are not accretion-powered 
pulsars \citep{2008A&ARv..15..225M}. From their spins and spin-down rates, it is estimated that their magnetic 
fields are usually much higher than those of conventional radio pulsars. It is 
found that, if they are NSs, their primary source of energy must be magnetic in order to explain their 
observed luminosities \citep{1996ApJ...473..322T}. On the other hand, the same luminosities can be explained 
through rotation \citep{2012PASJ...64...56M} or high magnetic fields \citep{2016JCAP...05..007M} if they are 
WDs. Many such objects are associated with supernova remnants 
\citep{1997ApJ...486L.129V,2002ApJ...574L..29L,2004ApJ...609L..13K,2012ApJ...751...53A,2012ApJ...761...66S} 
and for some of them direct measurement of surface magnetic fields has already been made through the proton 
cyclotron resonance feature \citep{2013Natur.500..312T,2016MNRAS.456.4145R}, confirming those objects to be 
NSs. However, some SGRs and AXPs have neither a supernova remnant association nor a measured magnetic field 
from any direct observation. Hence, the nature of these objects, at least for some of them, is not yet clear. 
The spin periods of these SGRs and AXPs are found to be more than $1\,$s \citep{2014ApJS..212....6O} and so 
they cannot be detected by the LIGO/Virgo detectors. However, they could be detected by various proposed 
future space-based detectors, such as \textit{LISA}, \textit{TianQin}, \textit{ALIA}, \textit{BBO} and 
\textit{DECIGO}. Here we show how GWs, if detected by these detectors, can easily distinguish whether such 
sources are NSs or WDs.

A similar analysis for SGRs and AXPs was recently carried out by \cite{2020MNRAS.498.4426S}. They considered 
all detected SGRs and AXPs and found that \textit{BBO} and \textit{DECIGO} should be able to detect them 
within $1$ to $5\,$yrs of integration time if they are WDs, whereas no proposed detectors could detect them if 
they are NSs. However, as mentioned earlier, for most of these SGRs and AXPs, either they are found to be 
associated with supernova remnants or their magnetic field strengths have been measured directly, which 
readily implies that they may be NSs and not WDs. Hence considering all of them to be WDs does not seem to 
provide a reasonable interpretation. In this paper, we consider only those sources that are not yet confirmed 
to be NSs and self-consistently calculate the signal-to-noise ratio (SNR) and thereby the necessary 
observation time, for the proposed GW detectors to detect these sources. Moreover, \cite{2020MNRAS.498.4426S} 
assumed the magnetic fields inside these sources have a dipole-like configuration. However, 
\cite{2014MNRAS.437..675W} showed that the magnetic field inside a compact object is likely to have a dominant 
toroidal field owing to the action of an $\Omega-$dynamo at the time of its birth. So in this paper, we 
evaluate the SNR of GW signal considering the stellar configurations of these objects are mostly governed by 
the toroidal fields present in their interiors. We also show equivalent results for poloidally dominated 
objects for completeness.

There have also been extensive attempts to detect super-Chandrasekhar WDs directly. Over the past one and half 
decades more than a dozen of these massive WDs have been indirectly inferred from observations of 
over-luminous type~Ia supernovae (SNe~Ia), such as SN~2003fg \citep{2006Natur.443..308H}, SN~2006gz 
\citep{2007ApJ...669L..17H}, SN~2009dc 
\citep{2009ApJ...707L.118Y,2010ApJ...714.1209T,2011MNRAS.410..585S,2011MNRAS.412.2735T,2012ApJ...756..191K}, 
SN~2007if \citep{2010ApJ...713.1073S,2010ApJ...715.1338Y,2012ApJ...757...12S}, SN~2013cv 
\citep{2016ApJ...823..147C}.  However, not even a single such massive WD has been observed directly. It has 
been shown that high magnetic fields and rotation can increase the maximum mass of a WD significantly 
\citep{2013PhRvL.110g1102D,2015MNRAS.454..752S,2019MNRAS.490.2692K}. High magnetic fields mean the WDs are 
less thermally luminous \citep{2020MNRAS.496..894G}, so such highly magnetized WDs are difficult to detect in 
any of the electromagnetic (EM) surveys, such as SDSS, \textit{Kepler}, \textit{Gaia} etc. We calculated the 
various time-scales for radiation in a previous paper if a magnetized WD behaves like a pulsar 
\citep{2020ApJ...896...69K}. It is well known that a pulsar-like object (where the magnetic field axis and 
rotation axis are misaligned) can emit both EM dipole and gravitational quadrupole radiations. The time-scales 
for the emission of these radiations depend on the magnetic field geometry and its strength. We showed that a 
WD with a high poloidal field ($B_\mathrm{p} \gtrsim 10^{12}\,$G) cannot emit either radiation for a long time 
because the rotation axis quickly aligns with the magnetic axis and it no longer behaves like a pulsar 
\citep{2020ApJ...896...69K}. In contrast, a WD with a weak surface poloidal magnetic field but a dominant 
toroidal field inside, can emit both EM dipole and gravitational quadrupole radiations for several years so 
that future space-based GW detectors will be able to detect the GWs. In addition, there are several WD 
pulsars, AR~Scorpii \citep{2016Natur.537..374M}, AE~Aquarii 
\citep{1994A&A...282..493R,1995MNRAS.275..649W,2004A&A...421.1131I}, RX~J$0648.0-4418$ 
\citep{2011ApJ...737...51M} etc. which could also emit gravitational radiation. However, some of these WD 
pulsars' exact masses are still unknown. Hence, we also estimate the time-scale for detection of 
super-Chandrasekhar WDs and WD pulsars along-with SGRs and AXPs through GWs and thereby discuss how to resolve 
the problems outlined above.

In Section~\ref{Sec: Modeling GW from a pulsar-like object}, we discuss the basic formalism of pulsar-like 
objects emitting GWs and the corresponding SNR to detect such objects by various GW detectors. Using these 
formulae, we calculate the GW strain and the corresponding SNR for various detectors, thereby discussing the 
time-scales for detecting the super-Chandrasekhar WDs in GW astronomy in Section~\ref{Sec: Possible detection 
of super-Chandrasekhar white dwarfs and white dwarf pulsars}. In this section, we also discuss the detection 
time-scale for some WD pulsars. Subsequently, in Section~\ref{Sec: Detection of SGR and AXP by GW detectors}, 
we calculate the SNR for those SGRs and AXPs which are not yet confirmed to be NSs. There we calculate the SNR 
separately considering them as WDs and NSs with different field configurations. We end with our conclusions in 
Section~\ref{Conclusions}.

\section{Modelling GWs from a pulsar-like object} \label{Sec: Modeling GW from a pulsar-like object}
\begin{figure}
	\centering
	\includegraphics[scale=0.4]{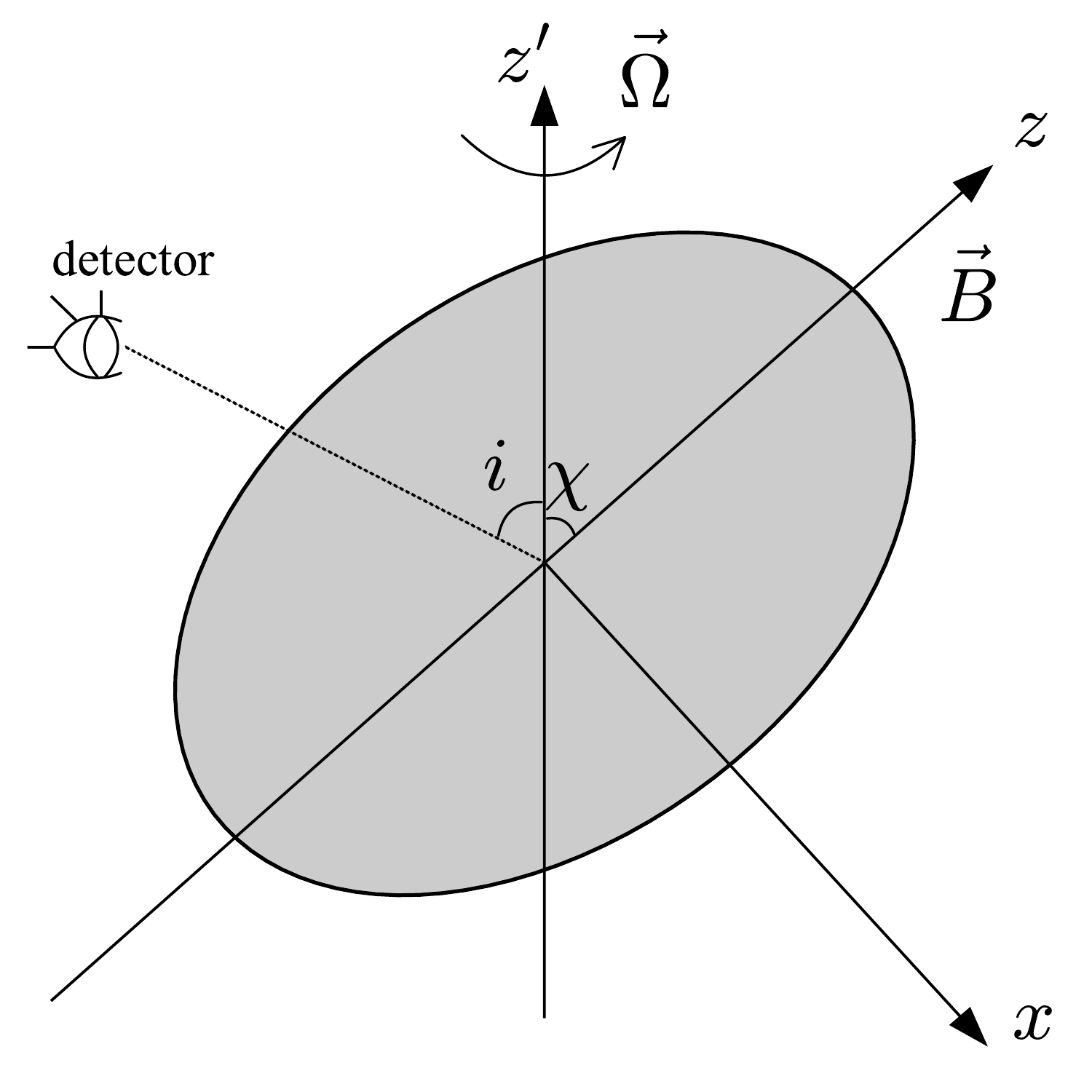}
	\caption{An illustrative diagram of a pulsar with magnetic field along the $z$-axis and rotation along the $z'$-axis. The angle $\chi$ is between these two axes and $i$ is the angle between the rotation axis and the observer's line of sight.}
	\label{Fig: pulsar}
\end{figure}
Our target is to explain pulsar-like objects. So we discuss our model of a pulsar, based on which we carry out 
the further calculations. Fig.~\ref{Fig: pulsar} shows a schematic diagram of a pulsar with $z$ being its 
magnetic field axis and $z'$ its rotational axis about which it has an angular velocity $\Omega$. The angle 
between these two axes is $\chi$ and the angle between the rotation axis and the observer's line of sight is 
$i$. It is already known that a toroidal magnetic field makes a star prolate 
\citep{2002PhRvD..66h4025C,2004ApJ...600..296I}, whereas a poloidal magnetic field, as well as rotation, 
deforms it to an oblate shape \citep{2008PhRvD..78d4045K,2012MNRAS.427.3406F}. Hence, the simultaneous 
presence of magnetic field and rotation, with a misalignment between the rotation and magnetic axes, makes a 
star a tri-axial system. Its moments of inertia about any three mutually perpendicular axes are different. 
Such a body can continuously emit significant EM dipole as well as gravitational quadrupole radiations. At a 
time $t$, the strain of the two polarizations of the GWs are given by \cite{1996A&A...312..675B} and 
\cite{maggiore} as
\begin{equation}\label{Eq: GW polarization_magnetic}
\begin{aligned}
h_+ &= A_{+,1}\cos\left({\Omega t}\right) + A_{+,2}\cos\left({2\Omega t}\right), \\
h_\times &= A_{\times,1}\sin\left({\Omega t}\right) + A_{\times,2}\sin\left({2\Omega t}\right),
\end{aligned}
\end{equation}
where
\begin{equation}\label{Eq: GW polarization_magnetic amplitude}
\begin{aligned}
A_{+,1} &= h_0 \sin 2\chi \sin i \cos i, \\
A_{+,2} &= 2h_0 \sin^2\chi (1 + \cos^2 i), \\
A_{\times,1} &= h_0 \sin 2\chi \sin i, \\
A_{\times,2} &= 4h_0 \sin^2\chi \cos i,
\end{aligned}
\end{equation}
with 
\begin{equation}\label{Eq: h0 magnetic}
h_0 = \frac{G}{c^4}\frac{\Omega^2(I_{zz}-I_{xx})}{d},
\end{equation}
where $G$ is Newton’s gravitational constant, $c$ is the speed of light, $d$ is the distance between the 
detector and the source, and $I_{xx}$ and $I_{zz}$ are the moments of inertia of the star about $x$- and 
$z$-axes respectively. Owing to the inclination $i$ the gravitational radiation is not isotropic. It is 
evident that a pulsar-like object can continuously emit GWs at two frequencies, $\Omega$ and $2\Omega$. This 
is another important consequence which is missing in the analysis of \cite{2020MNRAS.498.4426S}.

To obtain the compact object structure, we use a publicly available code named {\sc xns} (version 3.0), 
primarily aimed at the study of the axisymmetric structure of the NSs 
\citep{2014MNRAS.439.3541P}\footnote{\url{http://www.arcetri.astro.it/science/ahead/XNS/code.html}}. We 
appropriately tune this code for the WDs. Its advantage is that it can handle purely toroidal and purely 
poloidal magnetic field configurations for both uniformly and differentially rotating compact objects. 
However, we need to supply the equation of state (EoS) in a polytropic form, $\mathcal{P} = K \rho^\Gamma$, 
where $\mathcal{P}$ is the pressure, $\rho$ is the density, $\Gamma$ is the polytropic index and $K$ is the 
proportionality constant. In the case of a WD, we obtain $K$ and $\Gamma$ by fitting Chandrasekhar's 
well-known EoS \citep{1931ApJ....74...81C} for degenerate electron gas in various density intervals for a 
carbon-oxygen WD. However, the exact EoS of NSs is not known. So we assume $\Gamma = 2$ and $K = 1456\, \rm 
cm^5 \, g^{-1}\, s^{-2}$ according to \cite{2014MNRAS.439.3541P}. We obtain the ellipticity $\epsilon = 
\abs{I_{zz}-I_{xx}}/I_{xx}$ from the code by switching off the rotation and only incorporating the effect of 
the magnetic fields.

Because pulsar-like objects can emit both EM and gravitational radiations, they have associated dipole and 
quadrupole luminosities. The dipole luminosity for an axisymmetric WD is given by 
\cite{2000MNRAS.313..217M} as
\begin{equation}
L_\text{D} = \frac{2B_\mathrm{p}^2 R_\mathrm{p}^6 \Omega^4}{c^3} \sin^2\chi ~F(x_0),
\end{equation}
where $x_0=R_0 \Omega/c$, $B_\mathrm{p}$ is the strength of the magnetic field at the pole, $R_\mathrm{p}$ is 
the stellar radius at the pole, $R_0$ is the average radius of the WD and $F(x_0)$ is defined by
\begin{equation}
F(x_0) = \frac{x_0^4}{5\left(x_0^6 - 3x_0^4 + 36\right)} + \frac{1}{3\left(x_0^2 + 1\right)}.
\end{equation}
Similarly, the quadrupolar GW luminosity is given by \cite{1979PhRvD..20..351Z} as
\begin{equation}
L_\text{GW} = \frac{2G}{5c^5} (I_{zz}-I_{xx})^2 \Omega^6 \sin^2\chi \left(1+15\sin^2\chi\right).
\end{equation}
Owing to the emission of this energy, $\Omega$ and $\chi$ decrease over time. The variations of $\Omega$ and 
$\chi$ with respect to $t$ are given by \cite{2000MNRAS.313..217M} as
\begin{align}\label{Eq: radiation1}
\dv{(\Omega I_{z'z'})}{t} &= -\frac{2G}{5c^5} \left(I_{zz}-I_{xx}\right)^2 \Omega^5 \sin^2\chi \left(1+15\sin^2\chi\right) \nonumber\\ &- \frac{2B_\mathrm{p}^2 R_\mathrm{p}^6 \Omega^3}{c^3}\sin^2\chi ~F(x_0)
\end{align}
and \cite{2020ApJ...896...69K},
\begin{align}\label{Eq: radiation2}
I_{z'z'} \dv{\chi}{t} &= -\frac{12G}{5c^5} \left(I_{zz}-I_{xx}\right)^2 \Omega^4 \sin^3\chi \cos\chi \nonumber \\ &- \frac{2B_\mathrm{p}^2 R_\mathrm{p}^6 \Omega^2}{c^3}\sin\chi \cos\chi ~F(x_0),
\end{align}
where $I_{z'z'}$ is the moment of inertia of the body about its $z'$-axis. Earlier \cite{2020ApJ...896...69K} 
solved the set of equations~\eqref{Eq: radiation1} and~\eqref{Eq: radiation2} simultaneously to obtain the 
time-scale over which a WD can radiate.

A pulsar-like object radiates GWs at two frequencies. When we observe such a GW signal, whose strength 
remains unchanged during the observation time $T$, the corresponding detector's cumulative SNR is given by 
\cite{1998PhRvD..58f3001J} and \cite{2010MNRAS.409.1705B} as
\begin{equation}\label{Eq: SNR}
\mathrm{S/N} = \sqrt{\mathrm{S/N}_{\Omega}^2 + \mathrm{S/N}_{2\Omega}^2}\, ,
\end{equation}
where
\begin{align}\label{Eq: SNR v11}
\langle \mathrm{S/N}_{\Omega}^2\rangle &= \frac{\sin^2\zeta}{100}\frac{h_0^2T\sin^2 2\chi}{S_\mathrm{n}(f)}
\end{align}
and
\begin{align}\label{Eq: SNR v12}
\langle \mathrm{S/N}_{2\Omega}^2\rangle &= \frac{4\sin^2\zeta}{25}\frac{h_0^2T\sin^4 \chi}{S_\mathrm{n}(2f)},
\end{align}
where $\zeta$ is the angle between the interferometer arms and $S_\mathrm{n}(f)$ 
is the detector's power spectral density (PSD) at the frequency $f$ with $\Omega=2\pi f$. The data for various 
detectors' PSDs are taken from \cite{2015CQGra..32a5014M} and 
\cite{2020PhRvD.102f3021H}\footnote{\url{http://gwplotter.com}}. For our calculations, because we mostly deal 
with space-based interferometers such as \textit{LISA}, we assume $\zeta = 60\degree$. Note that the average 
is over all possible angles including $i$ which determine the object's orientation with respect to the 
celestial sphere reference frame. Note also that for one year GW observations with space-based antennas, there 
is a possibility of changing the antenna pattern with respect to time and this may lead to a change in the SNR 
by a factor of two. Moreover, if the spin-down is fast so that $\Omega$ changes quite rapidly, such time 
integration needs to be carried out in time-stacks $T_\mathrm{stack}$ such that, in each stack, $\Omega$ 
remains nearly constant. Also, an incoherent search with a time-stacking technique is 
computationally efficient compared to the aforementioned coherent search for a long time even if the signal 
strength remains unchanged during the observation period \citep{2000PhRvD..61h2001B,2005PhRvD..72d2004C}. 
This stacking method can be used for an all-sky search for 
unknown pulsars \citep{2012JPhCS.354a2010L}. Hence, the total observation time $T$ is divided into 
$\mathcal{N}$ time-stacks such that $T = \mathcal{N}T_\mathrm{stack}$. Assuming $f$, $\chi$ and $h_0$ remain 
nearly constant over the entire observation period $T$, adding $\mathcal{N}$ such stacks, the new cumulative 
SNRs are given by \cite{maggiore} as
\begin{align}\label{Eq: SNR v21}
\langle \mathrm{S/N}_{\Omega}^2\rangle &= \frac{\sin^2\zeta}{100}\frac{h_0^2 \sqrt{\mathcal{N}} T_\mathrm{stack}\sin^2 2\chi}{S_\mathrm{n}(f)} = \frac{\sin^2\zeta}{100}\frac{h_0^2 T\sin^2 2\chi}{\sqrt{\mathcal{N}}S_\mathrm{n}(f)}
\end{align}
and
\begin{align}\label{Eq: SNR v22}
\langle \mathrm{S/N}_{2\Omega}^2\rangle &= \frac{4\sin^2\zeta}{25}\frac{h_0^2 \sqrt{\mathcal{N}} T_\mathrm{stack}\sin^4 \chi}{S_\mathrm{n}(2f)} = \frac{4\sin^2\zeta}{25}\frac{h_0^2 T\sin^4 \chi}{\sqrt{\mathcal{N}}S_\mathrm{n}(2f)}.
\end{align}
Note that, in such a stacking technique, the SNR reduces by a factor $\mathcal{N}^{1/4}$ compared to the 
continuous integration in a full coherent search. However, if either $\Omega$ or $\chi$ changes significantly with 
time, the SNR needs to be calculated coherently for the each individual stack and then added incoherently to obtain 
the cumulative SNR. In this paper, hereinafter for all the discussion, we use the stacking technique with 
$T_\mathrm{stack}=1\,$hr. Moreover, to detect a CGW signal, we need $\langle\mathrm{S/N}\rangle\gtrsim 5$ for more 
than $95\%$ detection efficiency \citep{2011MNRAS.415.1849P}.

\section{Possible detection of super-Chandrasekhar white dwarfs and white dwarf pulsars} \label{Sec: Possible detection of super-Chandrasekhar white dwarfs and white dwarf pulsars}

\begin{figure}
	\centering
	\subfigure[$B_\mathrm{p} = 8.9\times10^{11}$ G]{\includegraphics[scale=0.33]{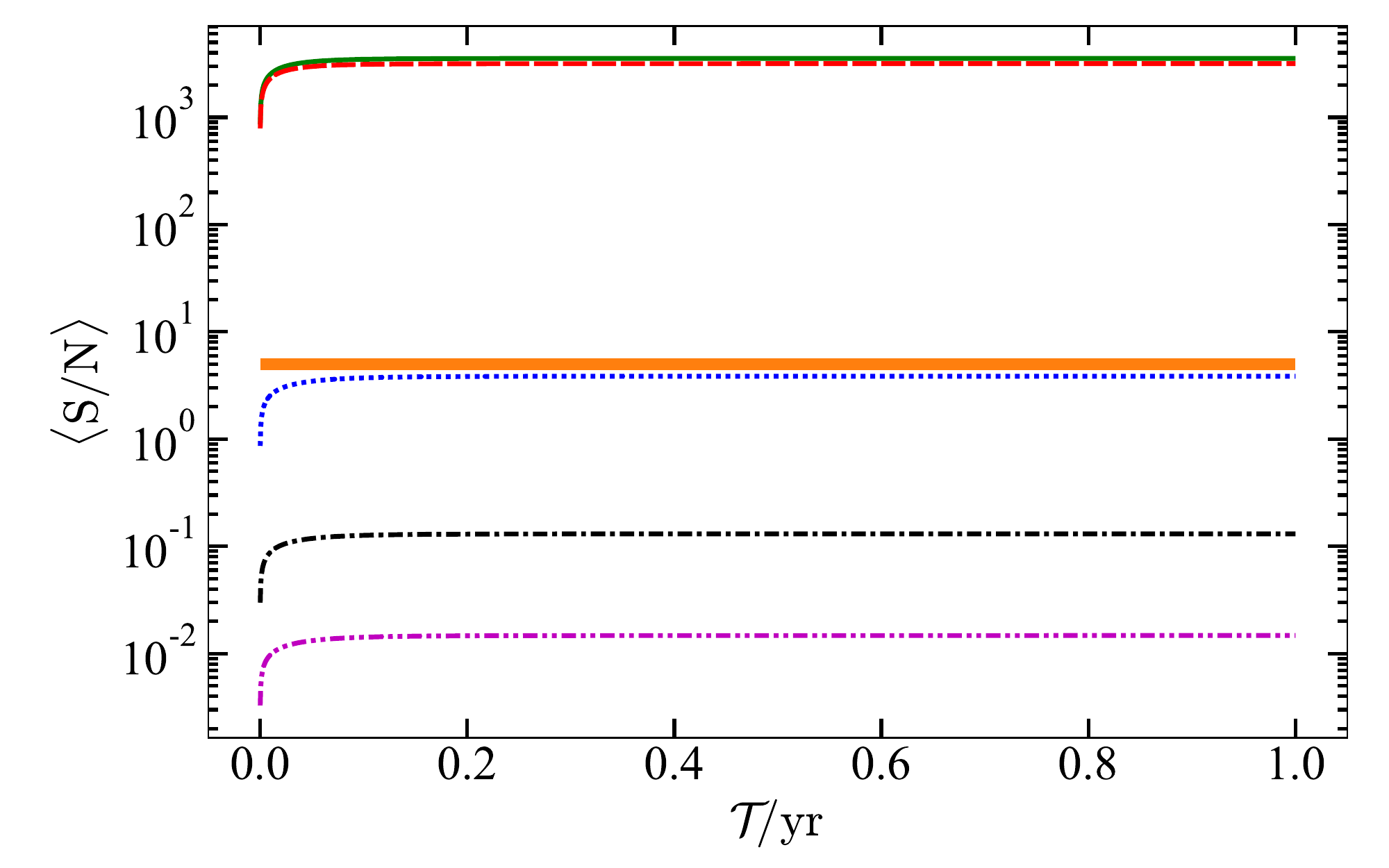}}
	\subfigure[$B_\mathrm{p} = 1.4\times10^{9}$ G]{\includegraphics[scale=0.33]{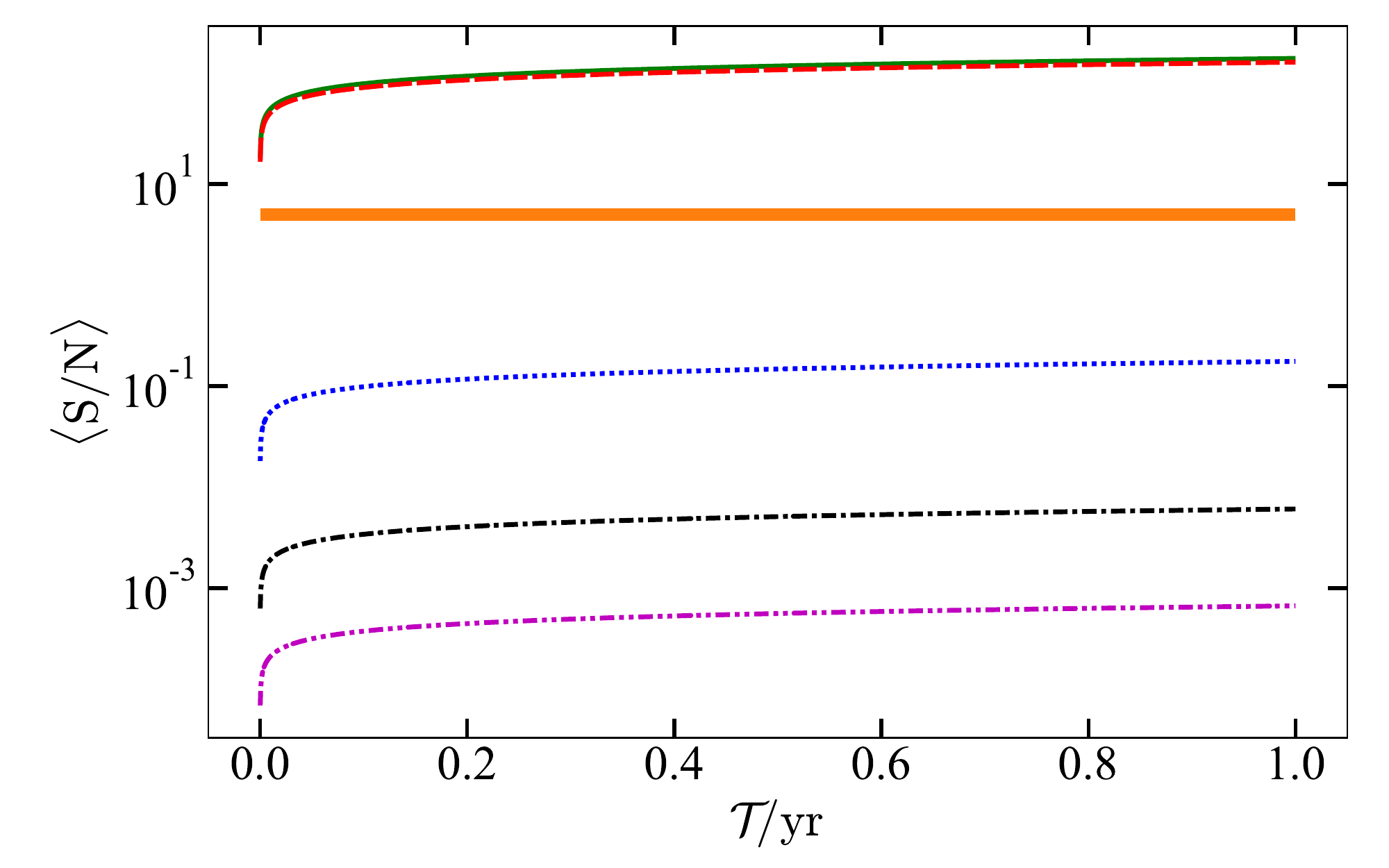}}
	\caption{SNR as a function of integration time for a poloidal magnetic field dominated WD with central 	density $\rho_\mathrm{c}=2\times 10^{10}\,$ g~cm$^{-3}$, spin period $P = 2\,$s, $\chi=30\degree$ and 	$d=100\,$pc. The solid green line represents \textit{BBO}, the dashed red line represents 	\textit{DECIGO}, the dotted blue line represents \textit{ALIA}, the dot-dashed black line represents 	\textit{TianQin} and the double-dot-dashed magenta line represents \textit{LISA}. The thick orange line corresponds to $\langle\mathrm{S/N}\rangle\approx5$.}
	\label{Fig: SNR poloidal field}
\end{figure}

\begin{figure}
	\centering
	\subfigure[$B_\mathrm{max} = 2.6\times10^{14}$ G]{\includegraphics[scale=0.33]{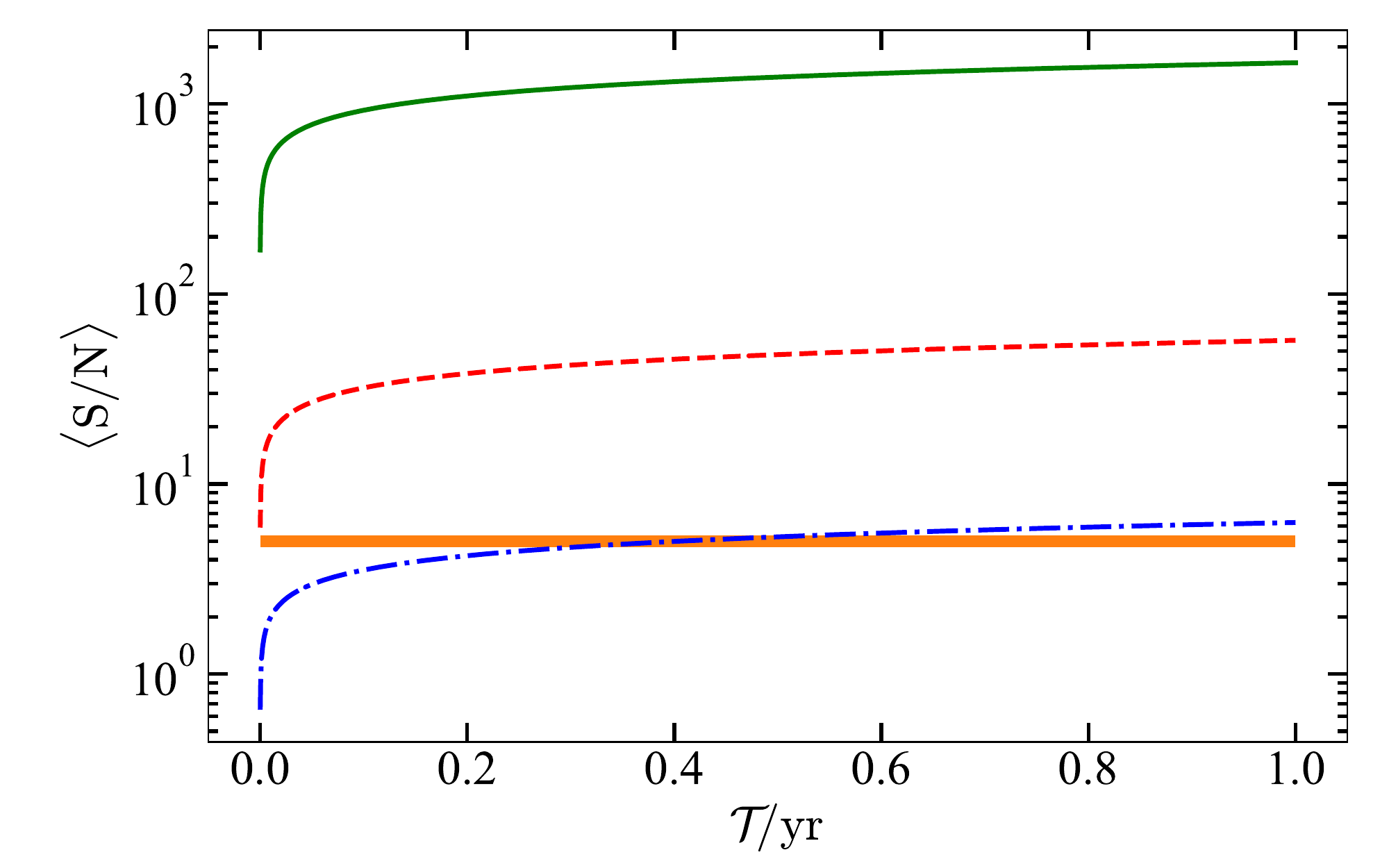}}
	\subfigure[$B_\mathrm{max} = 1.1\times10^{14}$ G]{\includegraphics[scale=0.33]{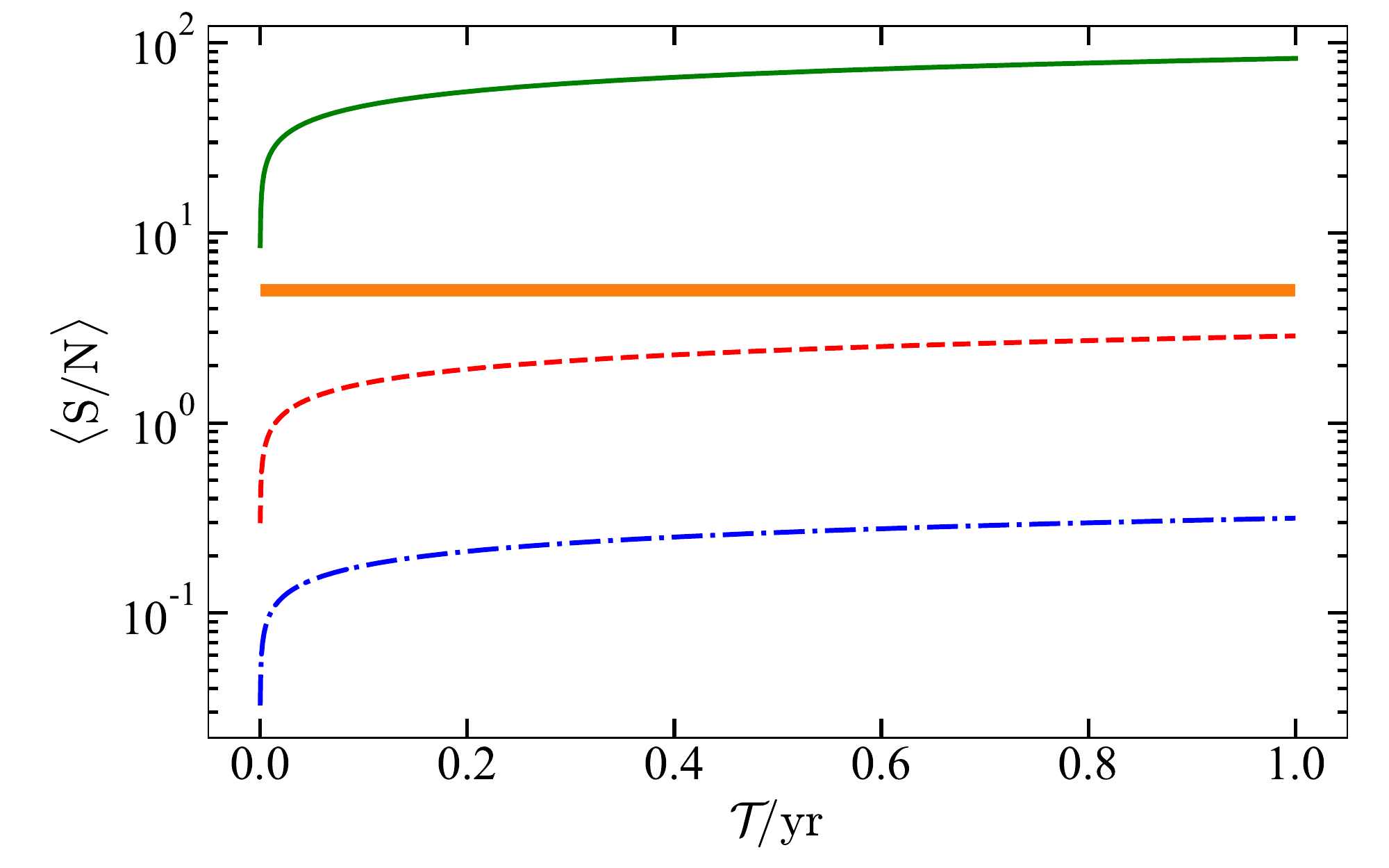}}
	\caption{As Fig.~\ref{Fig: SNR poloidal field} except for toroidal field dominated WDs. Here solid 	green line represents \textit{ALIA}, the dashed red line represents \textit{TianQin} and the dot-dashed blue line represents \textit{LISA}.}
	\label{Fig: SNR toroidal field}
\end{figure}

We first solve equations~\eqref{Eq: radiation1} and~\eqref{Eq: radiation2} simultaneously to obtain 
$\Omega(t)$ and $\chi(t)$ assuming poloidal field dominated WDs. Fig.~\ref{Fig: SNR poloidal field} shows the 
SNR as a function of time for poloidal field dominated WDs with different field strengths. In Fig.~\ref{Fig: 
SNR poloidal field}(a), the surface field is taken to be $B_\mathrm{p} \approx 8.9\times 10^{11}\,$G, which is 
decreased to $1.4\times10^9\,$G for Fig.~\ref{Fig: SNR poloidal field}(b). Because the surface field is strong 
in the first case $\Omega$ and $\chi$ decrease rapidly with time owing to the large $L_\text{D}$. 
It is found that the SNR increases for about one month and eventually saturates thereafter. This is 
because, in the stacking method, the power of the GW signal for each stack is added up. When $\Omega$ and $\chi$ 
decrease significantly, the strength of GW amplitude also decreases. This eventually results in the decrement 
of the power for later stacks. Hence, adding more stacks with significantly less power does not effectively 
change the cumulative SNR. On the other hand, when the magnetic field is lower 
(about $10^9\,$G), $L_\text{D}$ is lower and the SNR always increases with time for $1\,$yr because both 
$\chi$ and $\Omega$ remain nearly constant over this integration time, as depicted in Fig.~\ref{Fig: SNR 
poloidal field}(b). It is found that \textit{BBO} and \textit{DECIGO} should be able to detect such WDs 
quickly, while \textit{LISA}, \textit{TianQin} and \textit{ALIA} will cannot detect them.

Fig.~\ref{Fig: SNR toroidal field} shows the SNR as a function of time for toroidal field dominated WDs with 
different field strengths. In the {\sc xns} code it is not possible to choose a suitable toroidally dominated 
mixed field configuration. So we assume these toroidal dominated WDs have a poloidal surface field which is 
nearly four orders of magnitude smaller than the maximum toroidal field $B_\text{max}$ inside the WD. This is 
because, from the {\sc xns} code, we find that, for a purely poloidal field configuration in a WD, the centre 
may have a field strength two orders of magnitude higher than the surface poloidal field. Again, 
\cite{2014MNRAS.437..675W} showed that, inside a WD, the maximum toroidal field can be nearly two orders of 
magnitude higher than the maximum poloidal field. So we can assume that the maximum interior toroidal field 
can be nearly four orders of magnitude higher than the surface poloidal field. Of course, such a poloidal 
field cannot change the shape and size of the WD as does the toroidal field. So, owing to the limitations of 
the code, we run it for purely toroidal magnetic fields to obtain the shape and size of the WD. Moreover, the 
surface field strength is relatively very small (as is the dipole luminosity) and so it hardly changes 
$\Omega$ and $\chi$ within a $1\,$yr period. Fig.~\ref{Fig: SNR toroidal field}(a) shows the SNR for a WD with 
$B_\text{max}=2.6\times10^{14}\,$G with mass $1.7\,\rm M_\odot$. All the GW detectors except 
\textit{LISA} can easily detect such a WD almost immediately and \textit{LISA} can detect it in 
$5\,$months of integration. In contrast, when the field strength decreases ($B_\text{max}\approx 10^{14}\,$G) 
the SNR decreases and \textit{LISA} and \textit{TianQin} could no longer detect them as shown in \ref{Fig: SNR 
toroidal field}(b). However, they can still be detected by \textit{ALIA}, \textit{BBO} and \textit{DECIGO} 
within $1\,$yr of integration time.

\subsection{Detection of WD pulsars}
\begin{table}
	\centering
	\caption{Observational properties of WD pulsars.}
	\label{Table: WD pulsar}
	\begin{tabular}{|l|l|l|l|l|l|l|}
		\hline
		Name & $P$/s & $\dot{P}$/$\rm s\,s^{-1}$ & $M$/$\rm M_\odot$ & $d$/pc \\
		\hline
		AE Aquarii & 33.08 & $5.64\times10^{-14}$ & $0.8-1.0$ & 100 \\
		RX J$0648.0-4418$ & 13.18 & $6.00\times10^{-15}$ & $1.23-1.33$ & 650 \\
		AR Scorpii & 118.2 & $3.92\times10^{-13}$ & $0.81-1.29$ & 110 \\
		\hline
	\end{tabular}
\\ \cite[References:][]{1994A&A...282..493R,1995MNRAS.275..649W,2004A&A...421.1131I,2011ApJ...737...51M,2016Natur.537..374M,2016ApJ...831L..10G}
\end{table}
\begin{figure}
	\centering
	\includegraphics[scale=0.33]{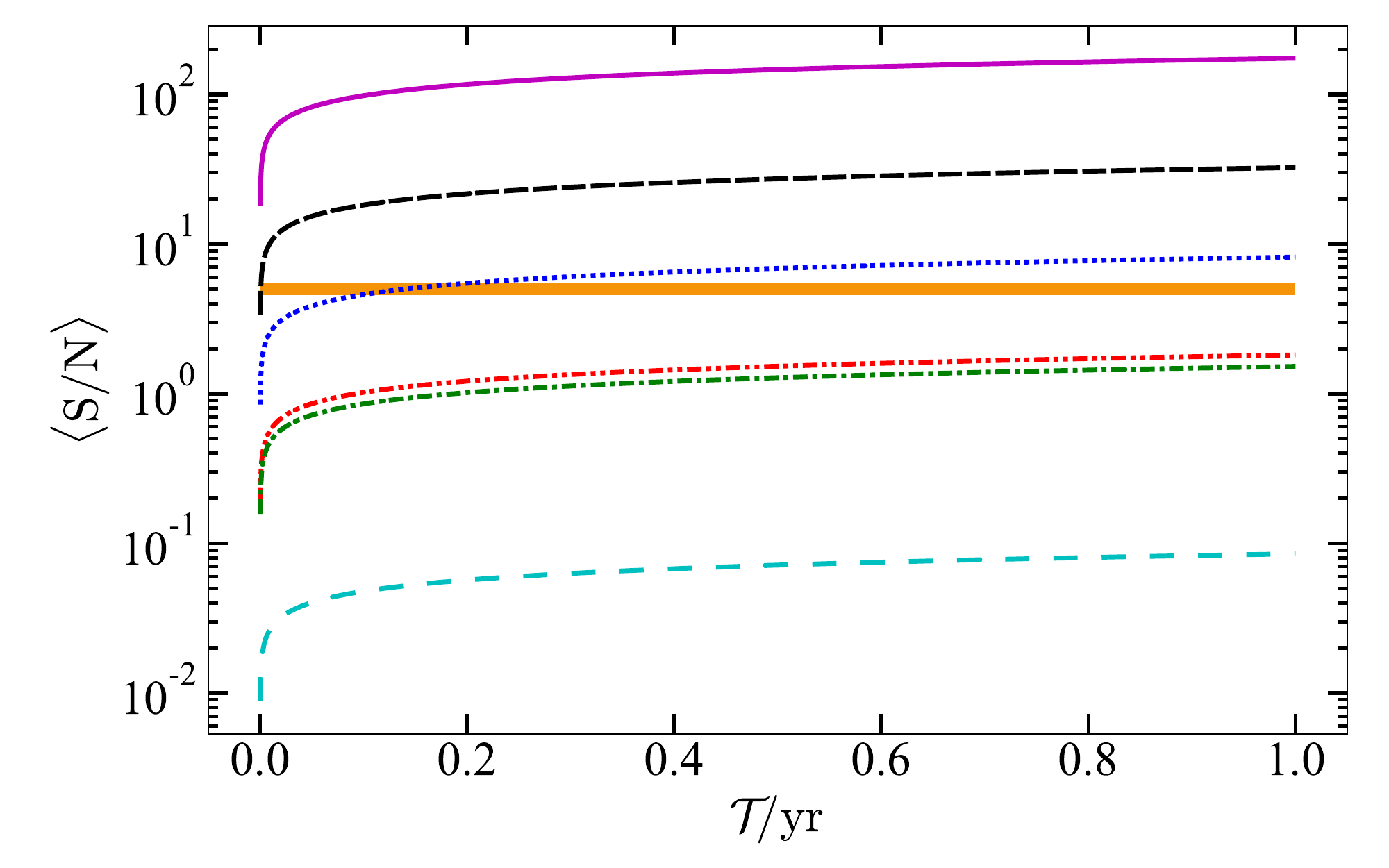}
	\caption{SNR as a function of integration time for AE~Aquarii with $M=0.9\,\rm M_\odot$ and $\chi=45\degree$. Here the solid magenta, dashed black and double-dot-dashed red lines show the SNR of \textit{BBO}, \textit{DECIGO} and \textit{ALIA} respectively when the source is a toroidally dominated WD, while the dotted blue, dot-dashed green and the loosely-dashed cyan lines represent the SNR in the case of a poloidally dominated WD. The thick orange line corresponds to $\langle\mathrm{S/N}\rangle\approx5$.}
	\label{Fig: AE aqr}
\end{figure}

\begin{figure}
	\centering
	\includegraphics[scale=0.33]{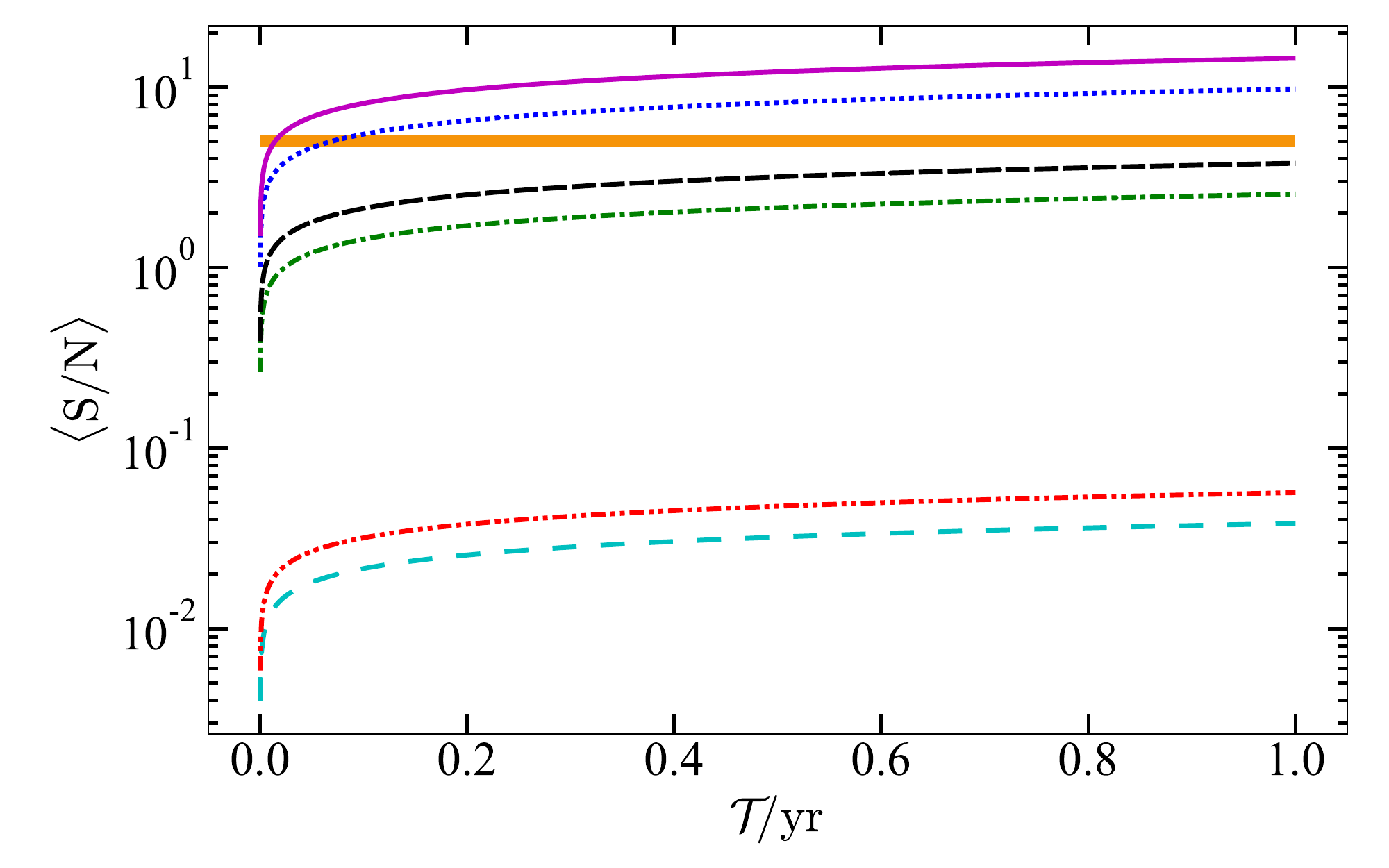}
	\caption{As Fig.~\ref{Fig: AE aqr} but for RX~J$0648.0-4418$ with $M=1.28\,\rm M_\odot$.}
	\label{Fig: RX J0648.0-4418}
\end{figure}

\begin{figure}
	\centering
	\subfigure[$M = 0.81\,\rm M_\odot$]{\includegraphics[scale=0.33]{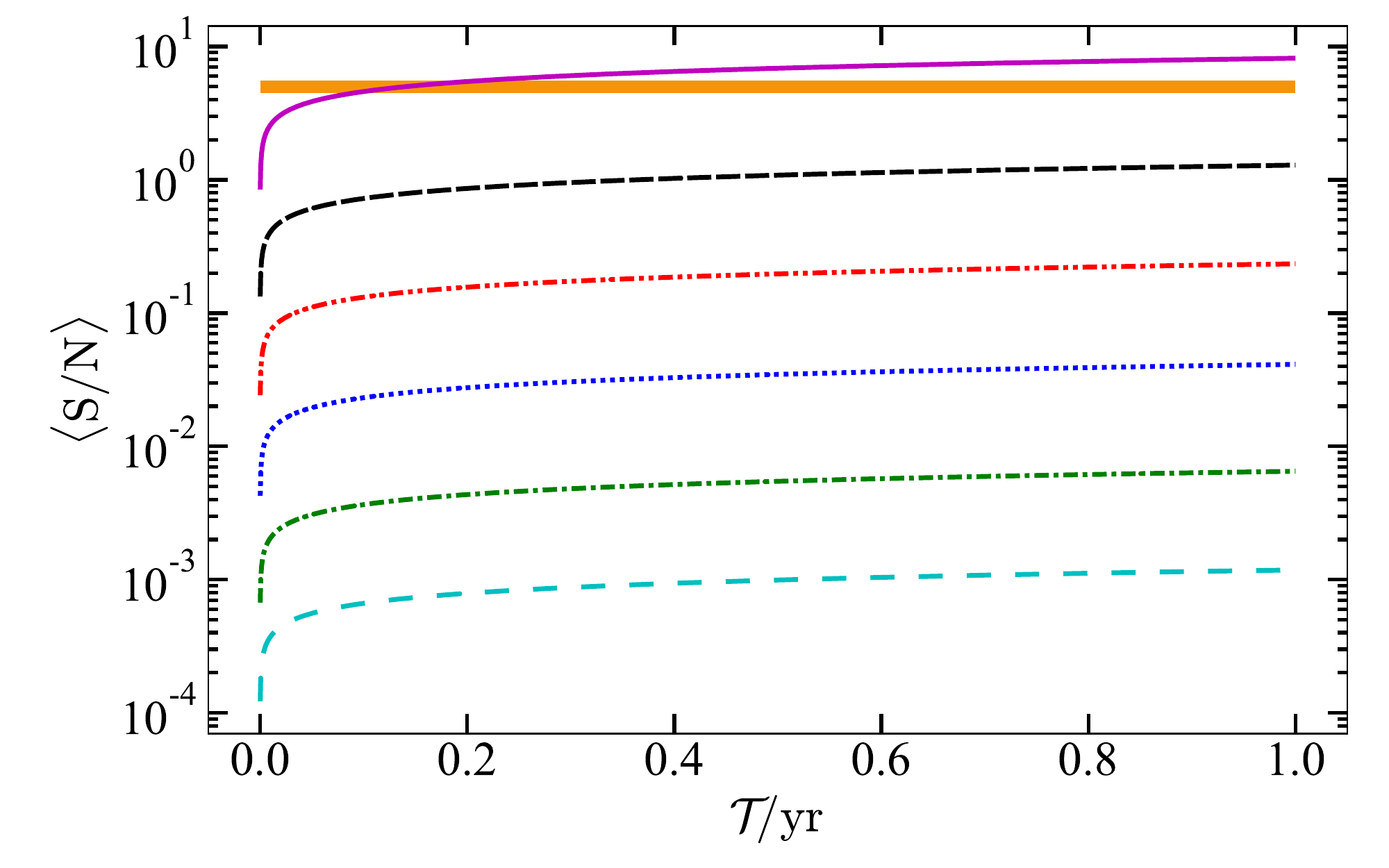}}
	\subfigure[$M = 1.29\,\rm M_\odot$]{\includegraphics[scale=0.33]{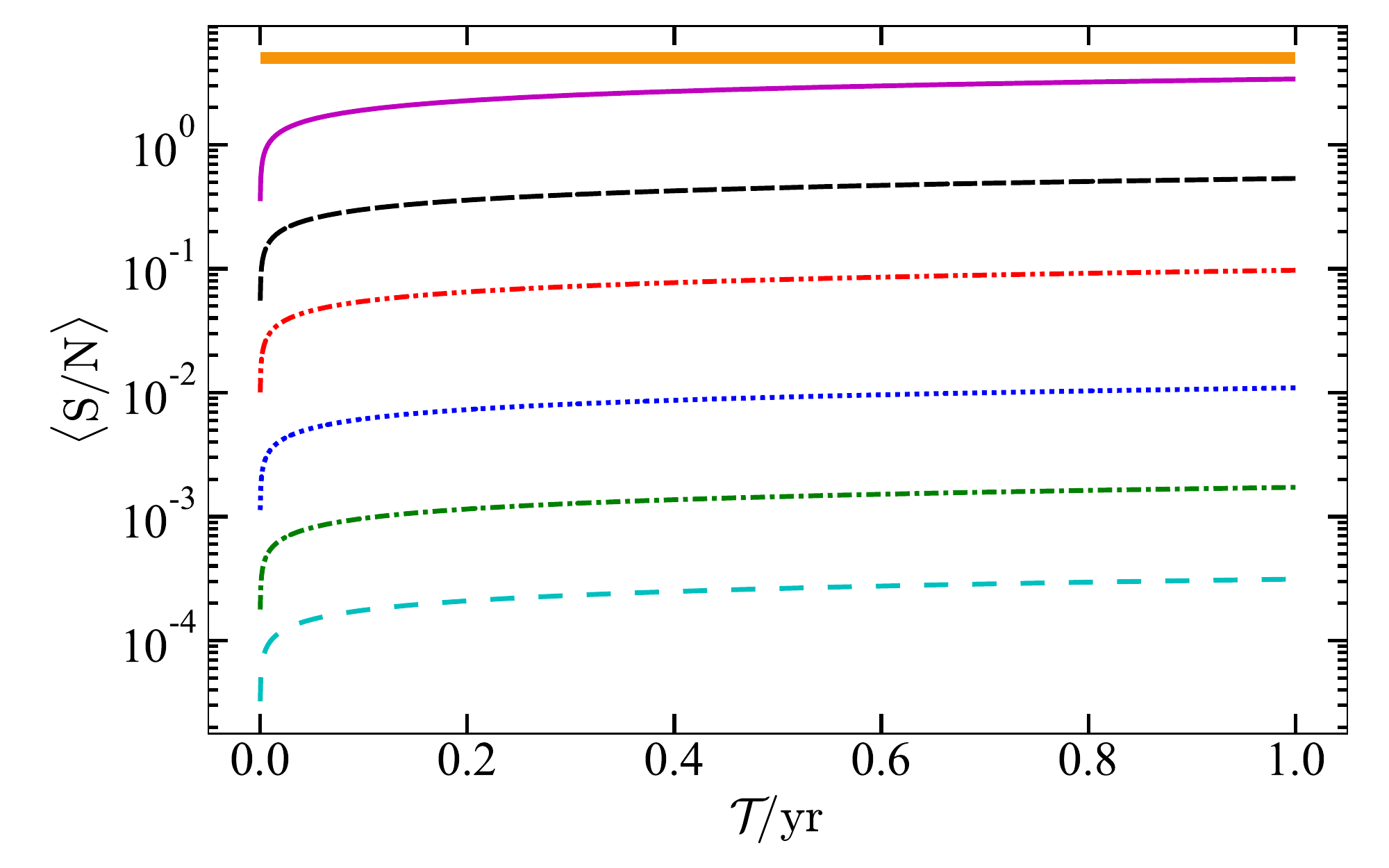}}
	\caption{As Fig.~\ref{Fig: AE aqr} but for AR Scorpii.}
	\label{Fig: AR Sco}
\end{figure}
As mentioned in the introduction, there are a few WD pulsars detected in EM surveys. Some of their 
observational properties are listed in Table~\ref{Table: WD pulsar}. Many of these WD pulsars are found with 
low-mass companions and are expected to be accreting from these binary partners. However, they differ from the 
intermediate polars (IPs) because their mass accretion rate is significantly smaller than that of the IPs, 
resulting in a lower X-ray luminosity \citep{2014MNRAS.442.2580P,2016Natur.537..374M}. Moreover, these WDs 
usually have a higher spin frequency than the other WDs. So their characteristic age ($P/2\dot{P}$) turns out 
to be between about $10^6$and $10^7\,$yr. These are WD pulsars, so they can emit both EM and GW radiations. 
Recently, \cite{2020MNRAS.492.5949S} attempted to explain the emission of GWs including the mass accretion 
from the binary companion as well as the magnetic deformation. However, their study lacked proper source 
modelling because they considered the magnetic field inside the WD to be a dipole only. As mentioned earlier, 
\cite{2014MNRAS.437..675W} showed that a toroidal field inside a WD can be much larger than its poloidal 
field. This may affect the deformation by the magnetic field and alter the amplitude of the GWs. Later, 
\cite{2017MNRAS.472.3564M} estimated the strength of GWs for highly magnetized WDs that have passed through a 
possible AR~Scorpii phase without considering a pure dipole field and assuming a mass accretion rate of about 
$10^{-8}\,\rm M_\odot\,$yr$^{-1}$. However, they did not estimate the AR~Scorpii's exact detection time-scale 
for any GW detector. Here we estimate the SNR for the GWs generated by the pulsation properties of these WD 
pulsars.

From equation~\eqref{Eq: radiation1}, if $P$ and $\dot{P}$ are known, $B_\mathrm{p}$ is given by
\begin{equation}\label{Eq: exact dipole magnetic field}
B_\mathrm{p} = \sqrt{\frac{c^3 I_{z'z'} P \dot{P}}{8 \pi^2 R_\mathrm{p}^6 \sin^2\chi ~F(x_0)} - \frac{4\pi^2 G}{5c^2 F(x_0)}\frac{\left(I_{zz}-I_{xx}\right)^2}{P^2 R_\mathrm{p}^6}\left(1+15\sin^2\chi\right)}.
\end{equation}
If $L_\text{D}\gg L_\text{GW}$, equation~\eqref{Eq: exact dipole magnetic field} can be approximated to
\begin{equation}
B_\mathrm{p} = \sqrt{\frac{c^3 I_{z'z'} P \dot{P}}{8 \pi^2 R_\mathrm{p}^6 \sin^2\chi ~F(x_0)}}.
\end{equation}
Moreover, for most of the WDs and NSs, $x_0\ll1$, which implies $F(x_0)\approx1/3$. Hence the above expression 
further reduces to the same form given by \cite{2016era..book.....C} as
\begin{equation}\label{Eq: dipole magnetic field}
B_\mathrm{p} = \sqrt{\frac{3c^3 I_{z'z'} P \dot{P}}{8 \pi^2 R_\mathrm{p}^6 \sin^2\chi}}.
\end{equation}
For all the WD pulsars, listed in Table~\ref{Table: WD pulsar}, $L_\text{D}$ is many orders of magnitude 
higher than $L_\text{GW}$. For example, for AE~Aquarii, we find $L_\text{D}\approx3\times 
10^{34}\rm\,erg\,s^{-1}$ and $L_\text{GW}\approx 10^{28}\rm\,erg\,s^{-1}$ if it is a poloidally dominated WD, 
whereas $L_\text{GW}\approx4\times 10^{30}\rm\,erg\,s^{-1}$ if it is a toroidally dominated WD. This implies 
that $L_\text{D}\gg L_\text{GW}$. So practically for this purpose, one can use either 
equation~\eqref{Eq: exact dipole magnetic field} or equation~\eqref{Eq: dipole magnetic field} without 
violating any physics. To detect these WD pulsars, $\chi$ must be non-zero and so we assume $\chi=45\degree$. 
Now, for each of the pulsars listed in Table~\ref{Table: WD pulsar}, the mass $M$ is known and we model them 
with the {\sc xns} code, assuming they are carbon-oxygen WDs, to obtain $R_\mathrm{p}$ and $I_{z'z'}$. If we 
assume that the WD has a dominant poloidal field, its shape and size are mostly determined by the poloidal 
field. Because the {\sc xns} code has the limitation that it can only model purely poloidal or purely toroidal 
magnetic fields along with rotation, we run it for a purely poloidal field to mimic the poloidally dominated 
WD. While for the toroidally dominated case, we assume the maximum toroidal magnetic field inside the WD is 
100~times larger than the maximum poloidal field as in the previous case and run the code with a purely 
toroidal field. \cite{2014MNRAS.437..675W} showed that such a ratio generally follows the stability criteria 
given by \cite{2009MNRAS.397..763B}. In this case, the shape and size are mostly determined by the internal 
toroidal field, while the surface dipole field contributes only to the dipole luminosity. Hence in this case, 
to obtain the size, we run the code assuming a purely toroidal field. Note that, owing to the small 
$\dot{P}$, $\Omega$ and $\chi$ remain nearly constant over a $1\,$yr period and so there is practically no need 
to solve equations~\eqref{Eq: radiation1} and \eqref{Eq: radiation2}. Hence we use equations~\eqref{Eq: SNR}, 
\eqref{Eq: SNR v21} and \eqref{Eq: SNR v22} to obtain the SNR for various GW detectors within $1\,$yr of 
integration time. We now discuss the detection time-scale for the WD pulsars as given below. Note that, 
because the positions of these pulsars are known, GW observations can be targeted searches.
\begin{itemize}
	\item AE~Aquarii: Fig.~\ref{Fig: AE aqr} shows the SNR for various detectors for the source AE~Aquarii 	
	assuming $M=0.9\,\rm M_\odot$. It is found that \textit{BBO} and \textit{DECIGO} can instantaneously 	
	detect this source if it is a toroidally dominated WD. If it has a dominant poloidal field only 	
	\textit{BBO} would be able to detect it within $2\,$months of integration time. Other detectors, such 	as 
	\textit{LISA} \textit{TianQin} or \textit{ALIA}, could detect it within $1\,$yr of integration.
	\item RX~J$0648.0-4418$: Fig.~\ref{Fig: RX J0648.0-4418} shows the SNR for RX~J$0648.0-4418$ with 	
	$M=1.28\,\rm M_\odot$. It is found that, apart from \textit{BBO}, no proposed GW detector would be able to 
	detect it within $1\,$yr of integration. \textit{BBO} would be able to detect it within $1\,$week or 
	$1\,$month, respectively, depending on whether it is a toroidally dominated or poloidally 
	dominated WD.
    \item AR~Scorpii: Because there is a larger mass range of $0.81$~to $1.29\,\rm M_\odot$ for AR~Scorpii, we 
    separately calculate the SNR for both extremes. Fig.~\ref{Fig: AR Sco}(a) shows the SNR for a mass of 
    $0.81\,\rm M_\odot$ and \ref{Fig: AR Sco}(b) for $1.29\,\rm M_\odot$. It is evident from both the figures 
    that only \textit{BBO} would be able to detect this source within $2\,$months of integration if it has a 
    mass of $0.81\,\rm M_\odot$ with a dominant toroidal field. No other proposed detector would be able to 
    detect it irrespective of its magnetic field configuration.
\end{itemize}

\section{Detection of SGRs and AXPs by GW detectors}\label{Sec: Detection of SGR and AXP by GW detectors}

\begin{table}
	\centering
	\caption{Observational properties of SGRs and AXPs.}
	\label{Table: SGR and AXP}
	\begin{tabular}{|l|l|l|l|l|l|l|}
		\hline
		Name & $P$/s & $\dot{P}$/$\rm s\,s^{-1}$ & $d$/kpc \\
		\hline
		1RXS J$170849.0-400910$ & 11.01 & $1.9\times10^{-11}$ & 3.8 \\
		3XMM J$185246.6+003317$ & 11.56 & $1.4\times10^{-13}$ & 7.1 \\
		4U $0142+61$ & 8.69 & $2.0\times10^{-12}$ & 3.6 \\
		SGR $1833-0832$ & 7.57 & $3.5\times10^{-12}$ & 2.0* \\
		XTE J$1810-197$ & 5.54 & $7.8\times10^{-12}$ & 3.5 \\
		\hline
	\end{tabular}
\\ *Actual distance is not known at this time.
\\ \url{http://www.physics.mcgill.ca/~pulsar/magnetar/main.html}
\end{table}

\begin{figure*}
	\centering
	\subfigure[Poloidal field dominated WD]{\includegraphics[scale=0.36]{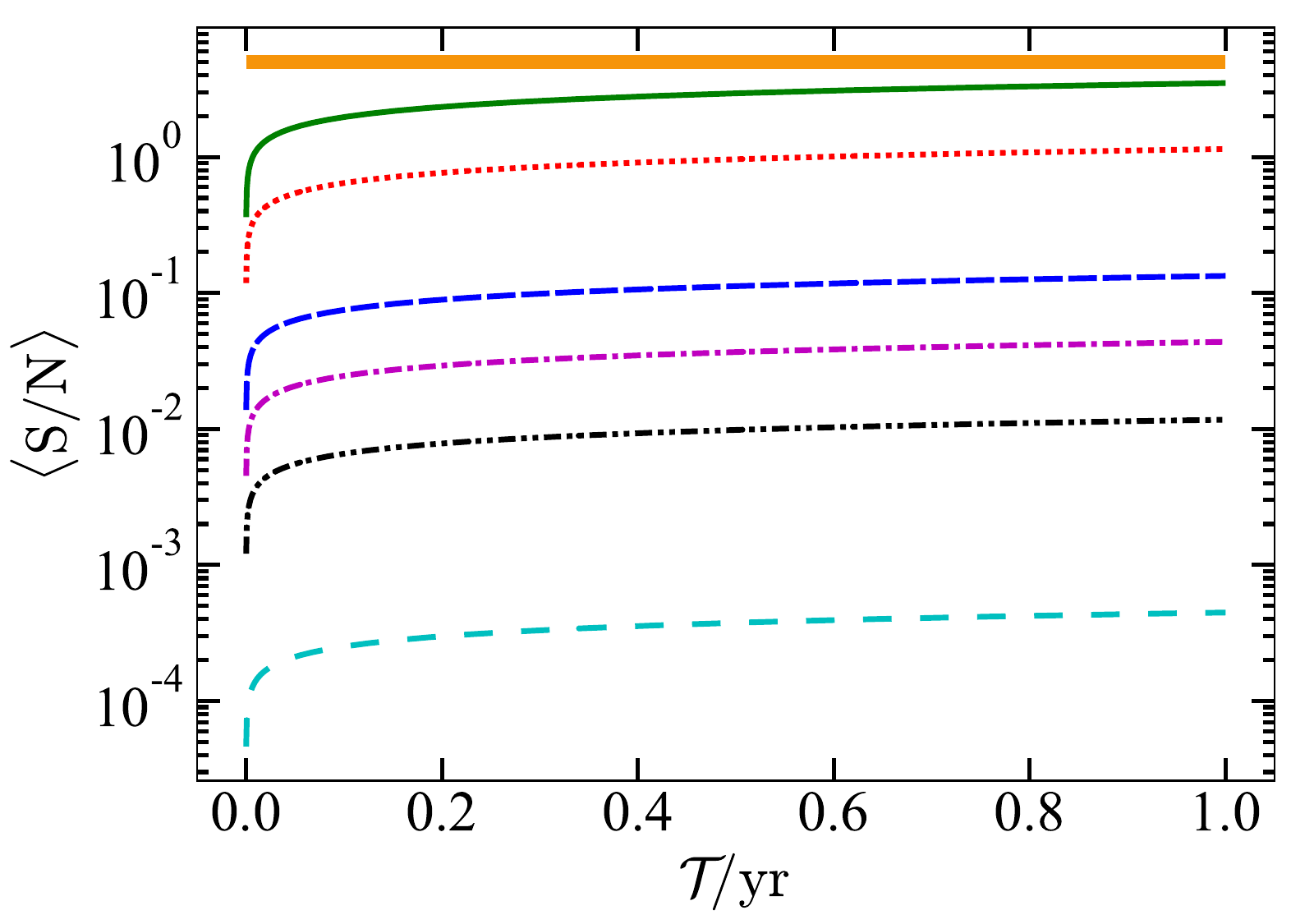}}
	\subfigure[Toroidal field dominated WD]{\includegraphics[scale=0.36]{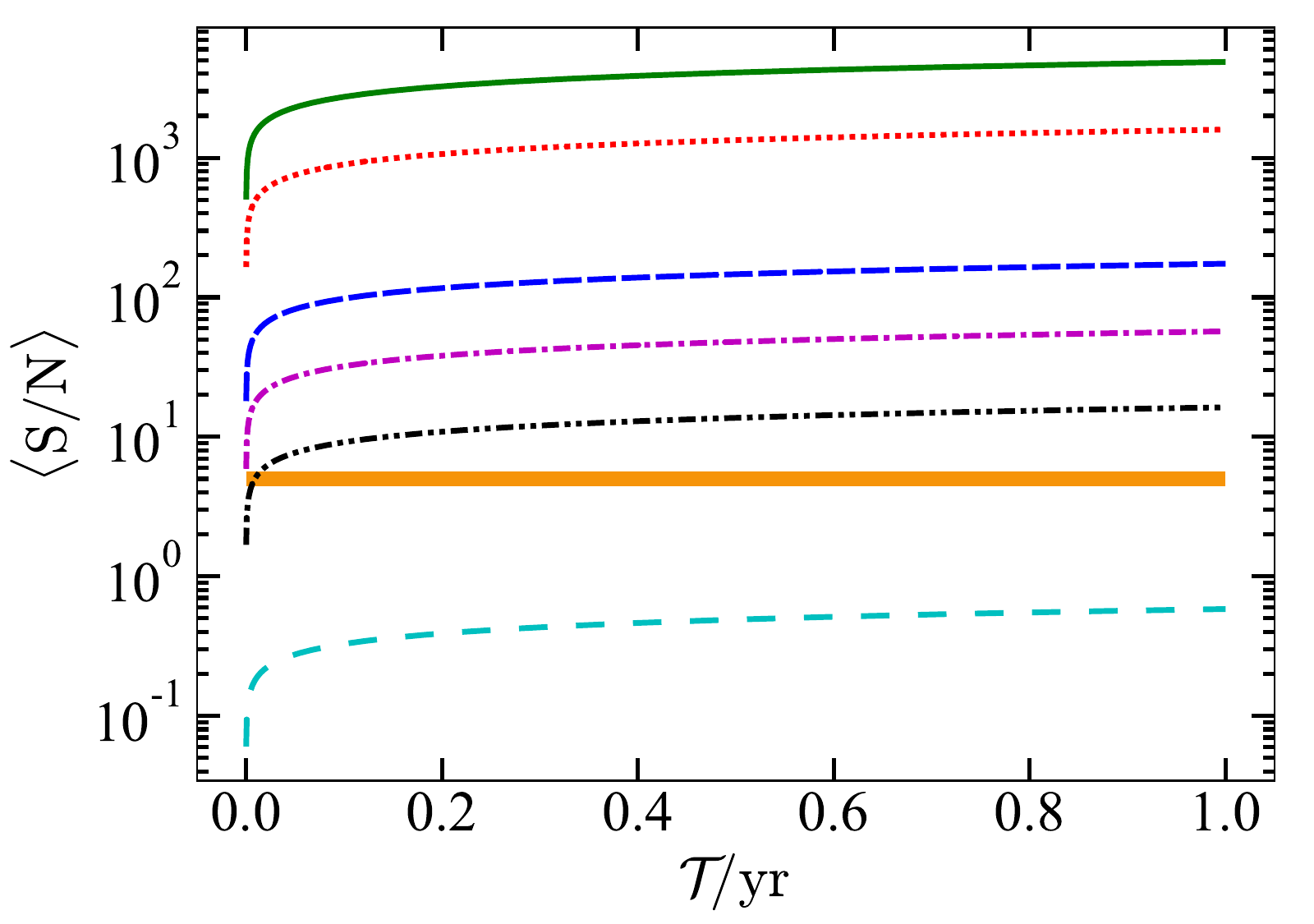}}
	\subfigure[NS with radius 14 km]{\includegraphics[scale=0.36]{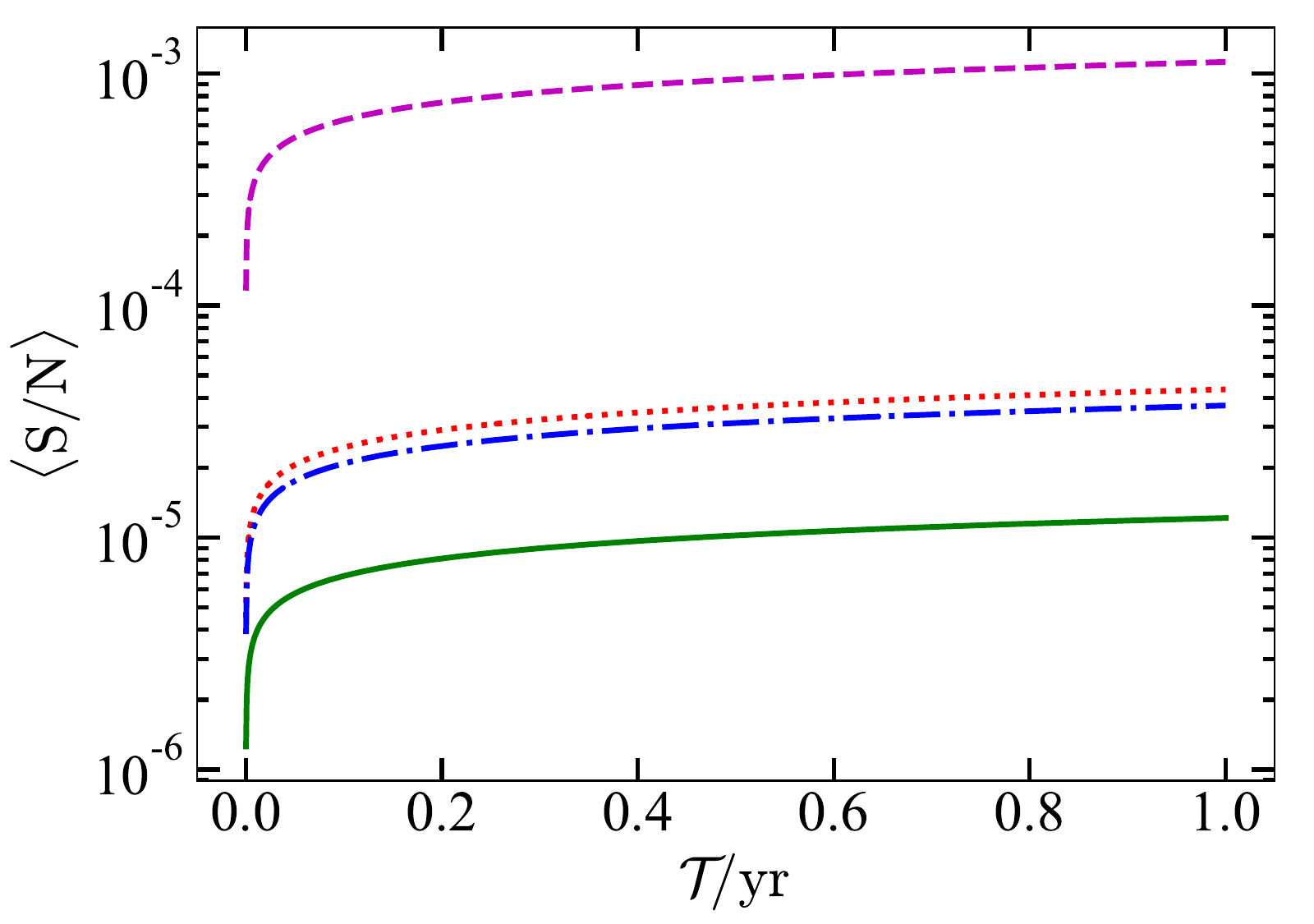}}
	\caption{SNR as a function of integration time for 1RXS~J$170849.0-400910$ assuming $\chi=45\degree$. For parts~(a) and~(b), solid green, dotted red and double-dot-dashed black lines correspond to a WD with radius $4000\,$km for \textit{BBO}, \textit{DECIGO} and \textit{ALIA} respectively, while dashed blue, dot-dashed magenta and loosely-dashed cyan lines represent a WD with radius $1000\,$km. For part~(c), solid green and dot-dashed blue lines correspond to poloidally dominated NSs for \textit{DECIGO} and \textit{BBO} respectively, whereas dotted red and dashed magenta lines represent toroidally dominated NSs. The thick orange line corresponds to $\langle\mathrm{S/N}\rangle\approx5$.}
	\label{Fig: 1RXS_J170849.0-400910}
\end{figure*}

\begin{figure*}
	\centering
	\subfigure[Poloidal field dominated WD]{\includegraphics[scale=0.36]{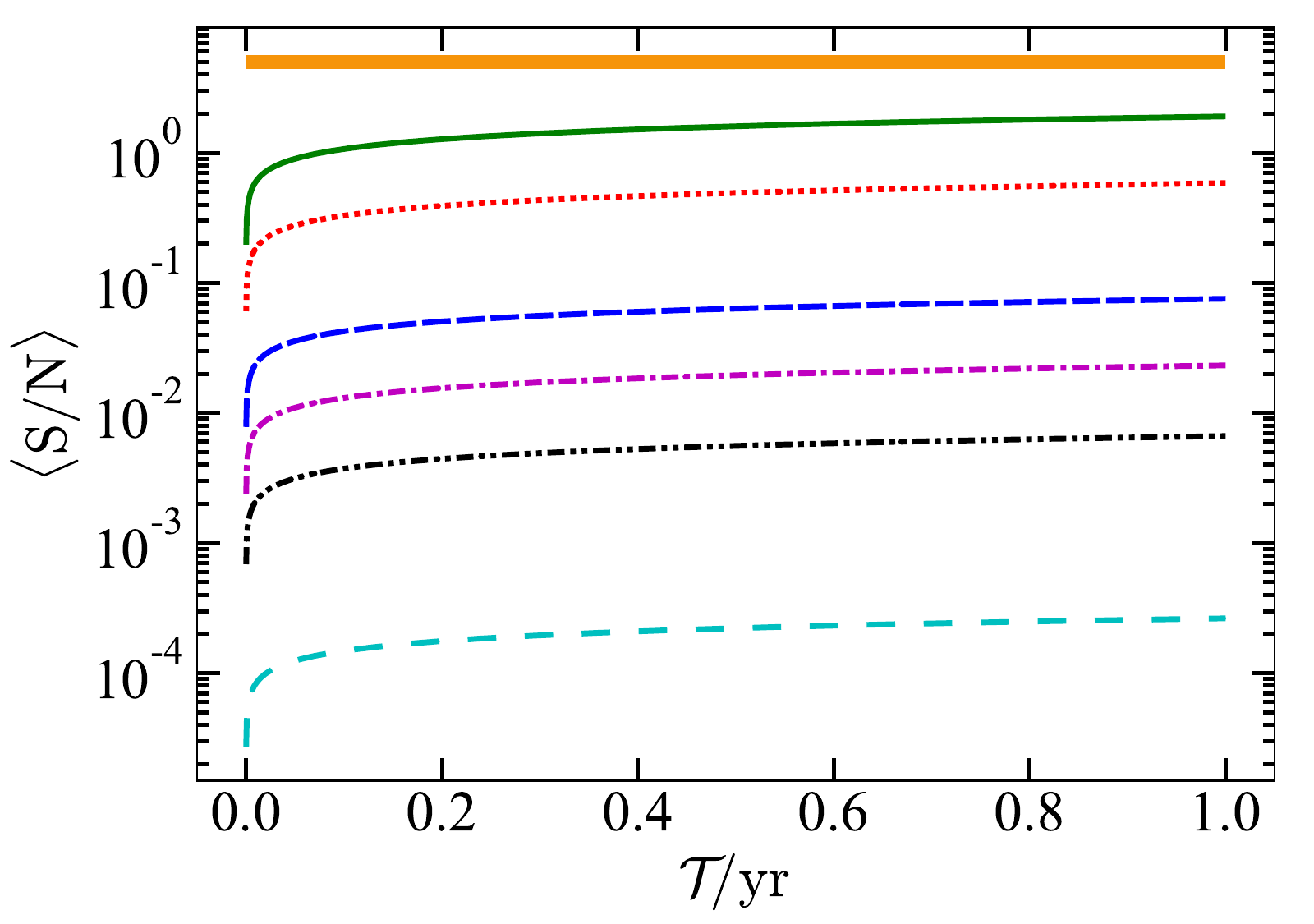}}
	\subfigure[Toroidal field dominated WD]{\includegraphics[scale=0.36]{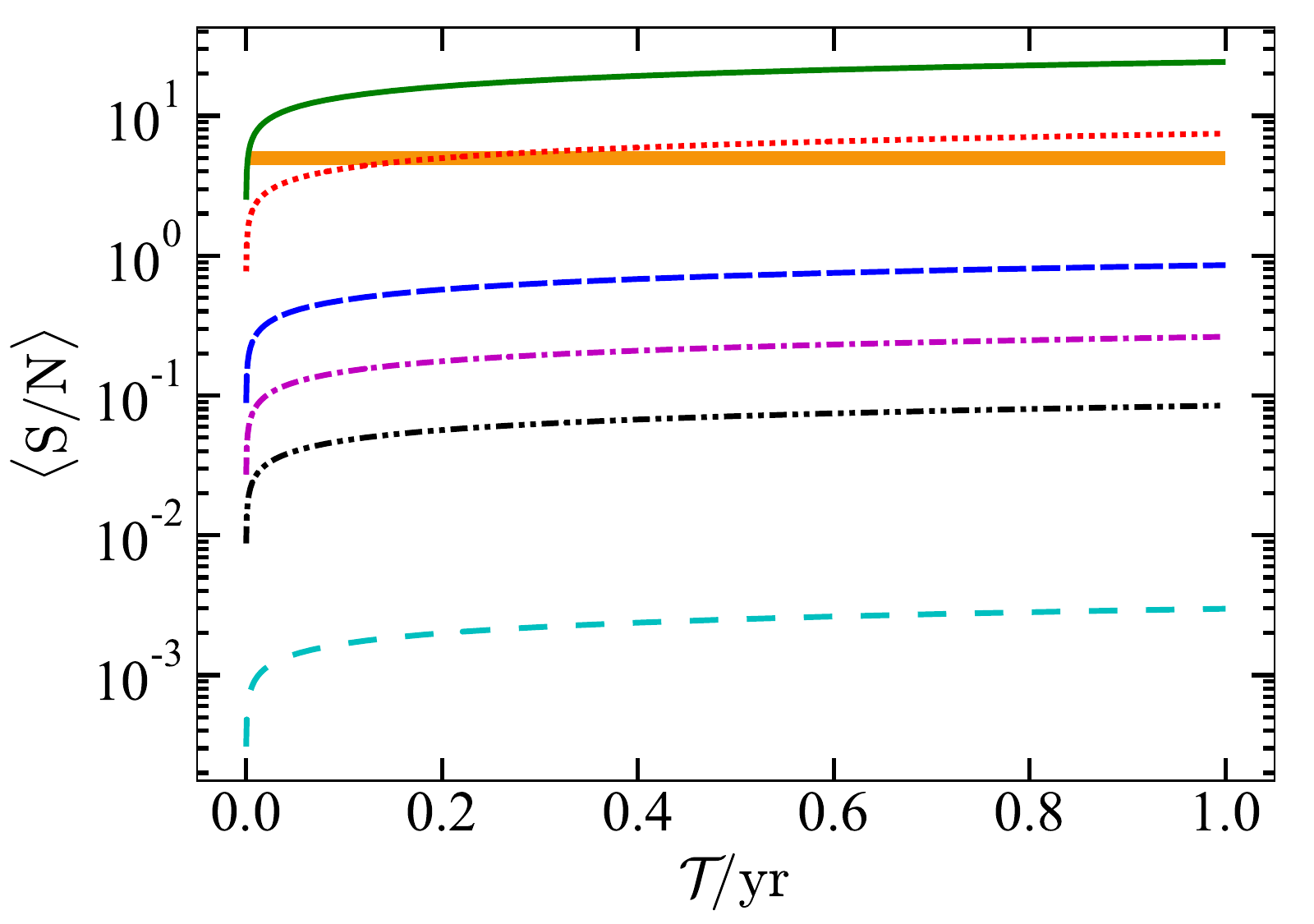}}
	\subfigure[NS with radius 14 km]{\includegraphics[scale=0.36]{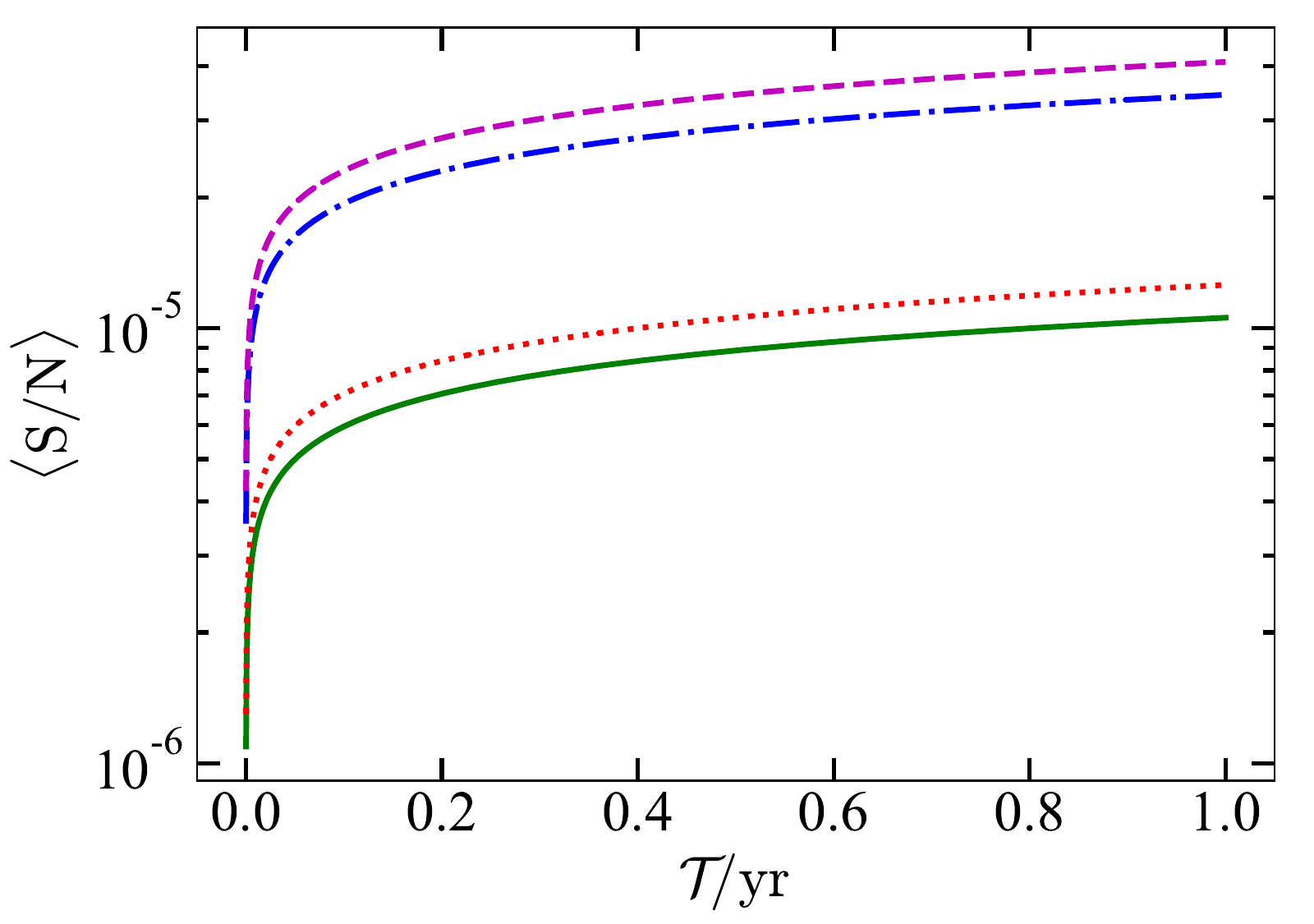}}
	\caption{As Fig.~\ref{Fig: 1RXS_J170849.0-400910} but for 3XMM~J$185246.6+003317$.}
	\label{Fig: 3XMM_J185246.6+003317}
\end{figure*}

\begin{figure*}
	\centering
	\subfigure[Poloidal field dominated WD]{\includegraphics[scale=0.36]{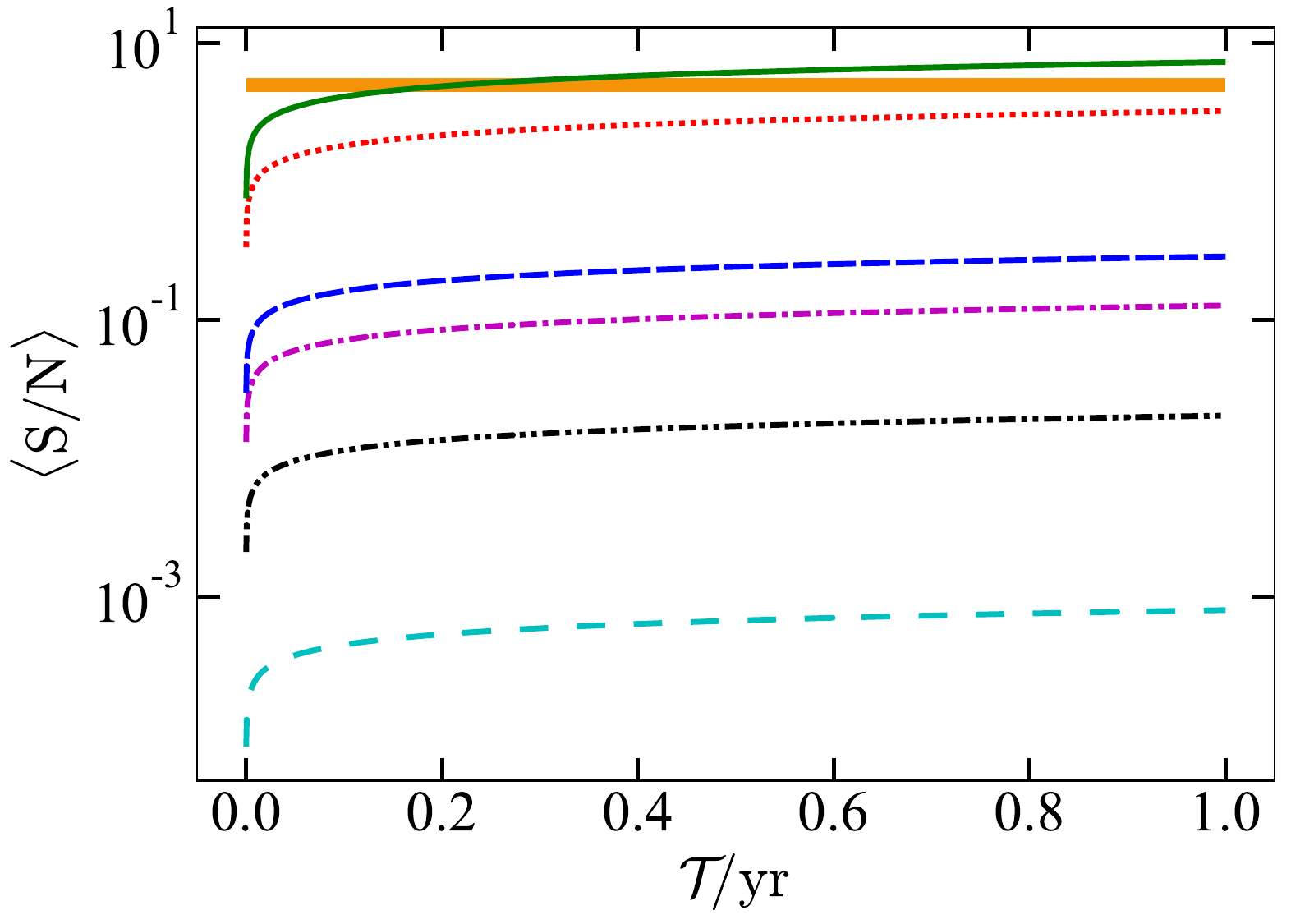}}
	\subfigure[Toroidal field dominated WD]{\includegraphics[scale=0.36]{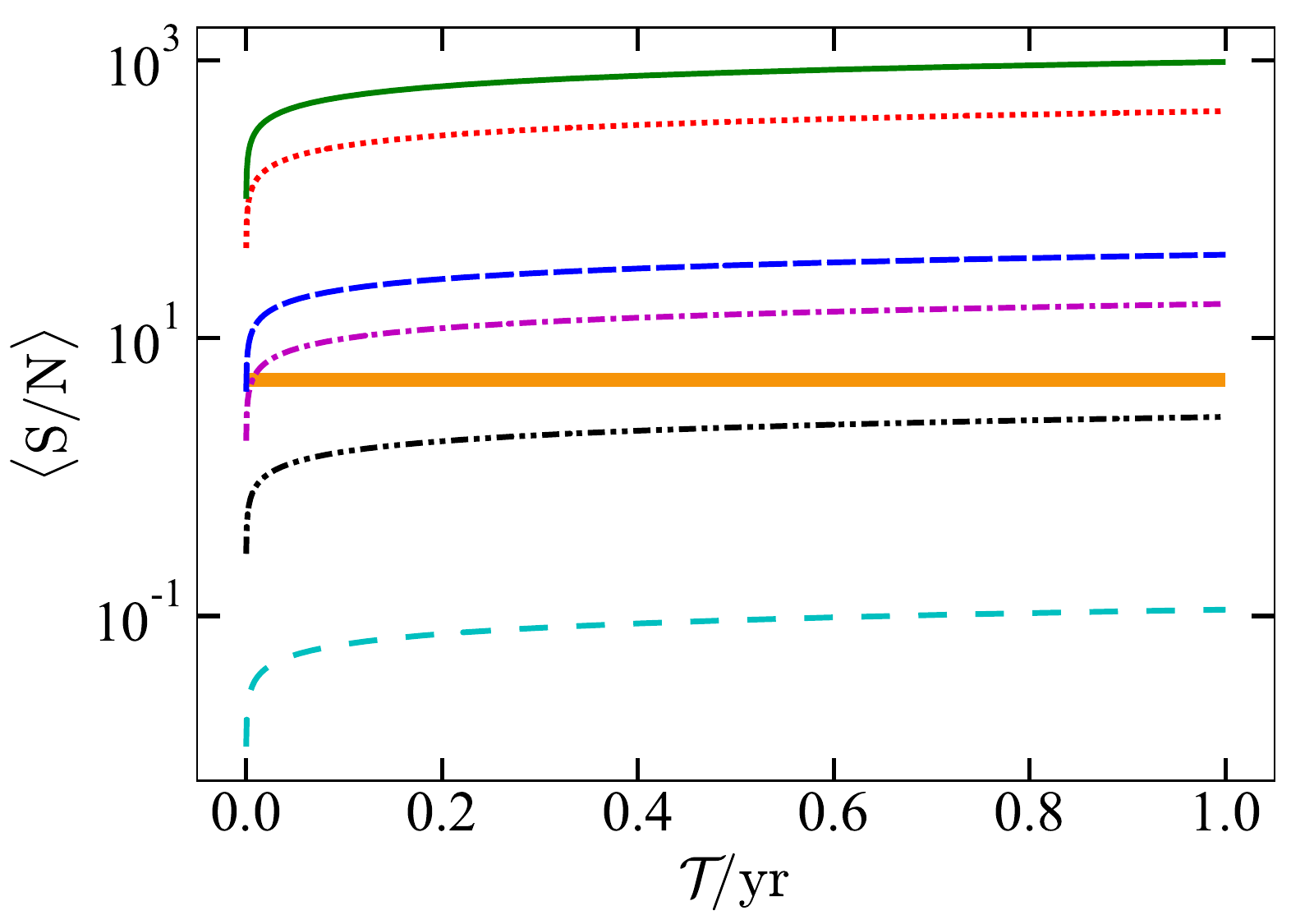}}
	\subfigure[NS with radius 14 km]{\includegraphics[scale=0.36]{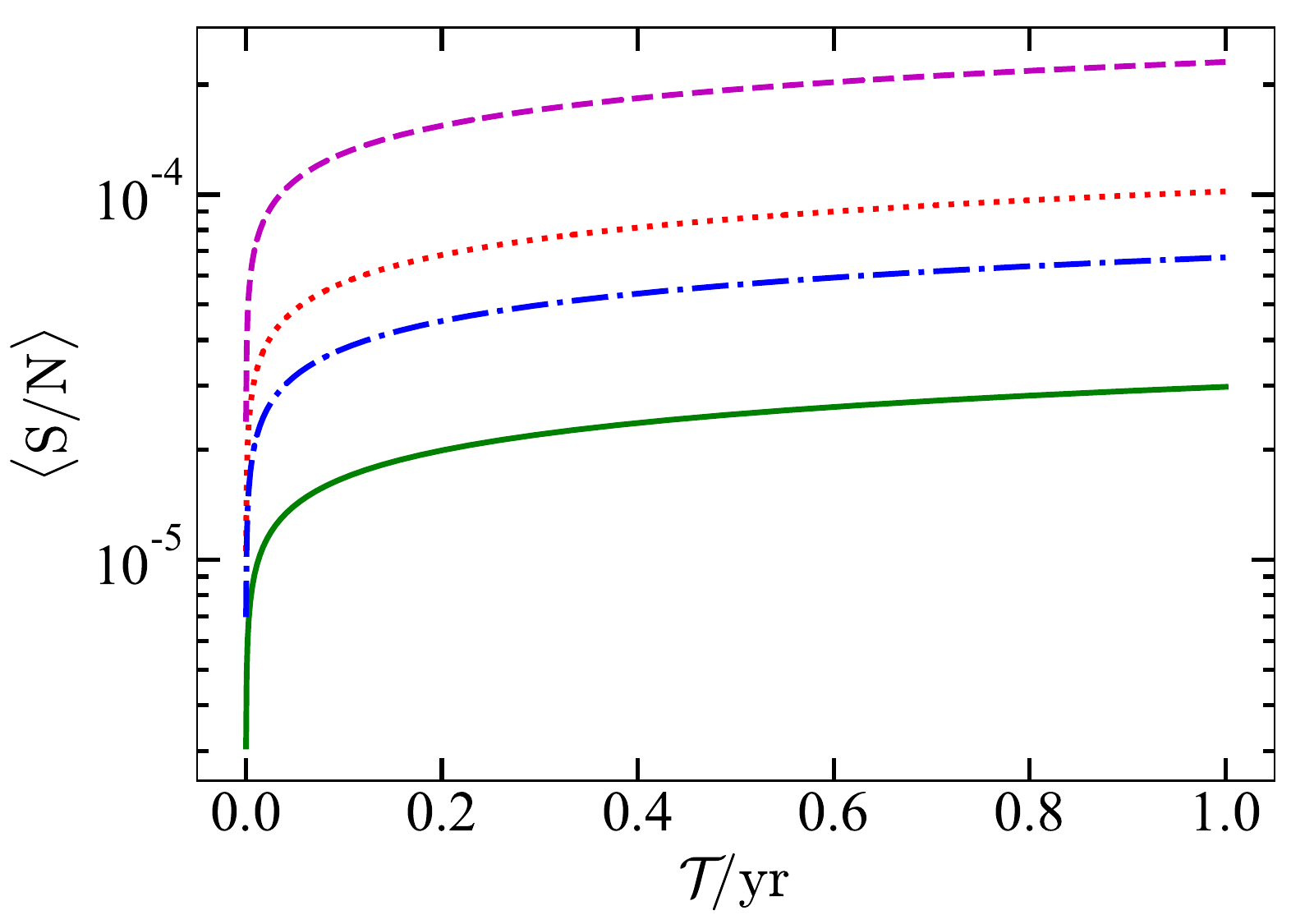}}
	\caption{As Fig.~\ref{Fig: 1RXS_J170849.0-400910} but for 4U~$0142+61$.}
	\label{Fig: 4U_0142+61}
\end{figure*}

\begin{figure*}
	\centering
	\subfigure[Poloidal field dominated WD]{\includegraphics[scale=0.36]{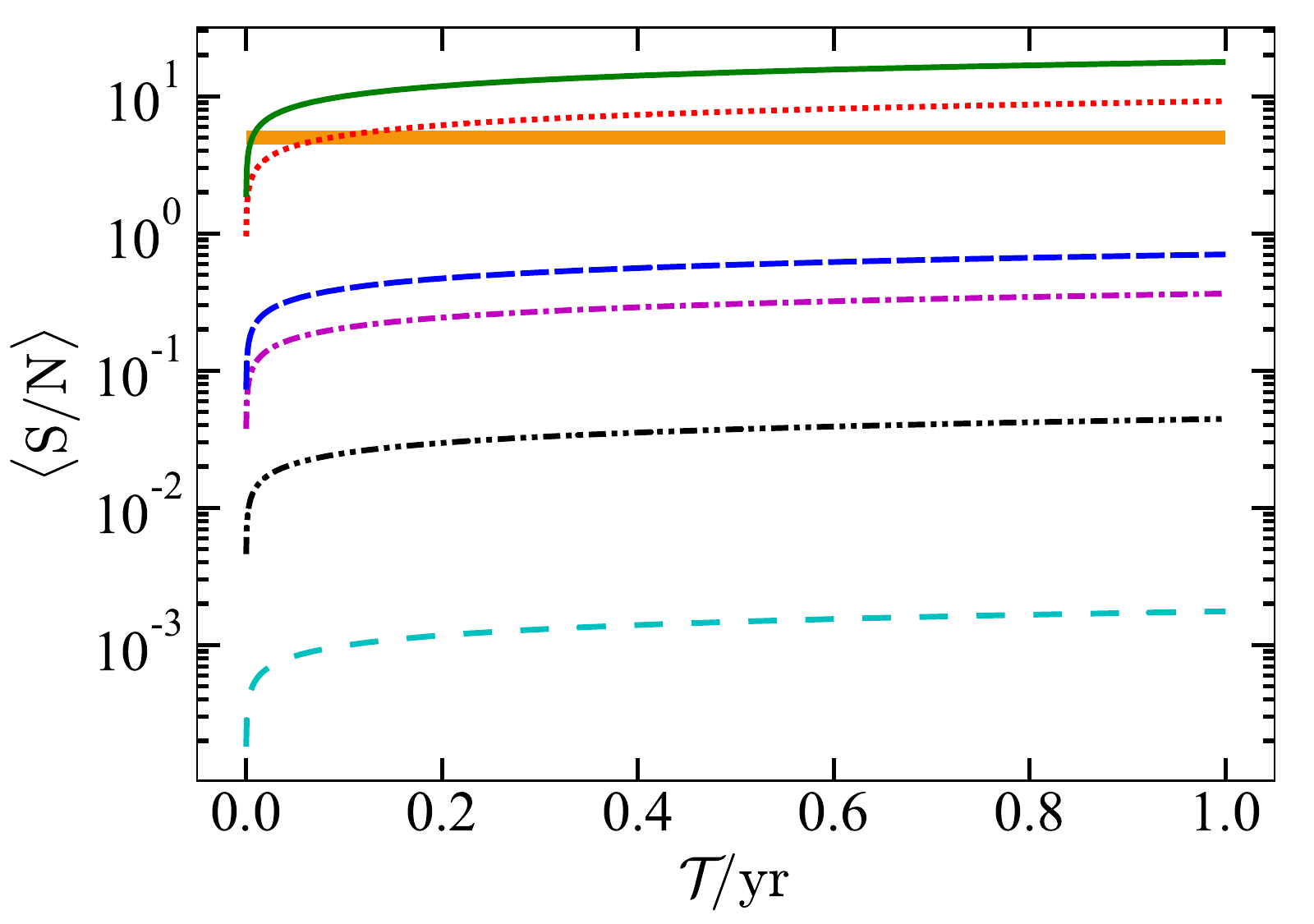}}
	\subfigure[Toroidal field dominated WD]{\includegraphics[scale=0.36]{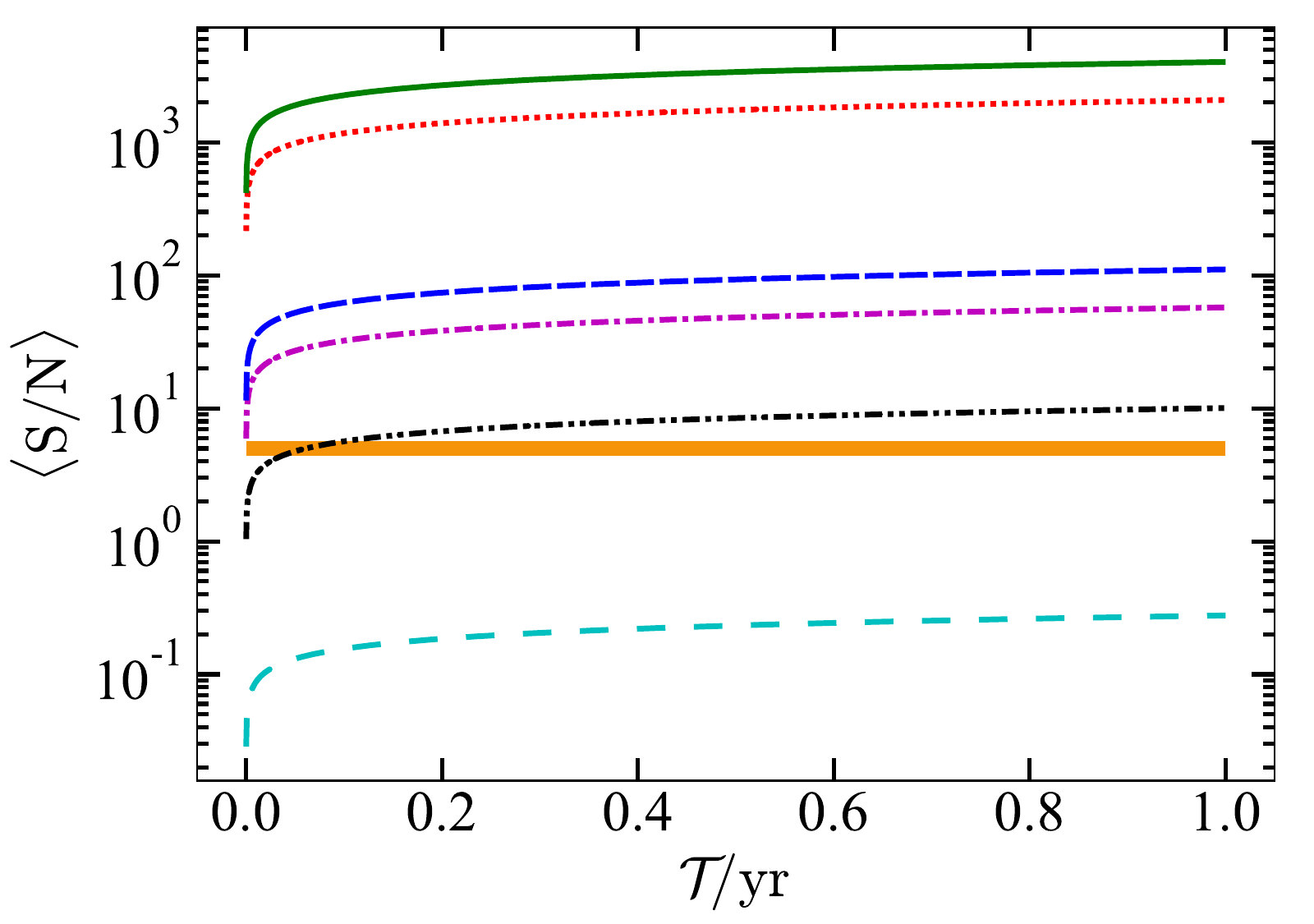}}
	\subfigure[NS with radius 14 km]{\includegraphics[scale=0.36]{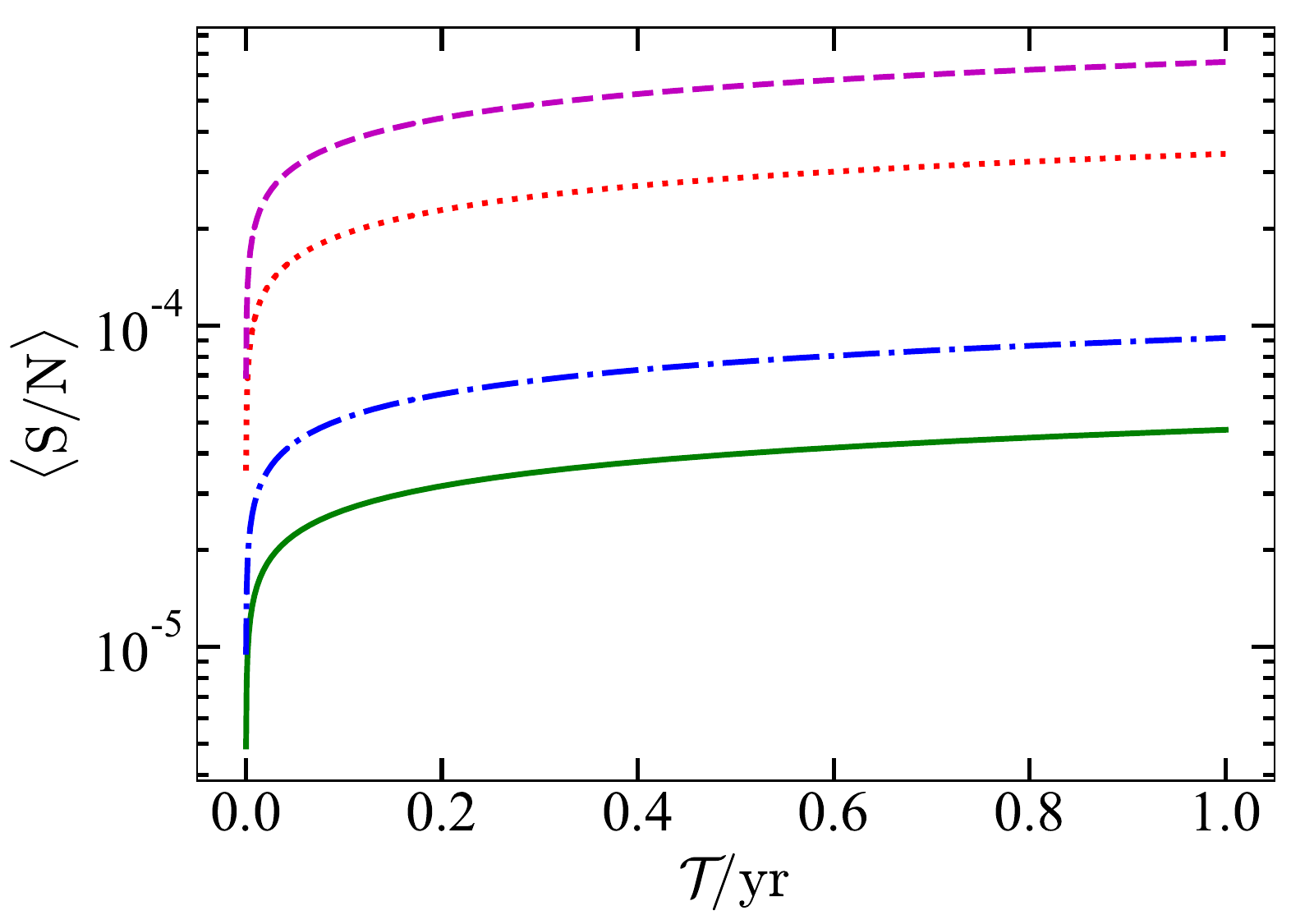}}
	\caption{As Fig.~\ref{Fig: 1RXS_J170849.0-400910} but for SGR~$1833-0832$.}
	\label{Fig: SGR_1833-0832}
\end{figure*}

\begin{figure*}
	\centering
	\subfigure[Poloidal field dominated WD]{\includegraphics[scale=0.36]{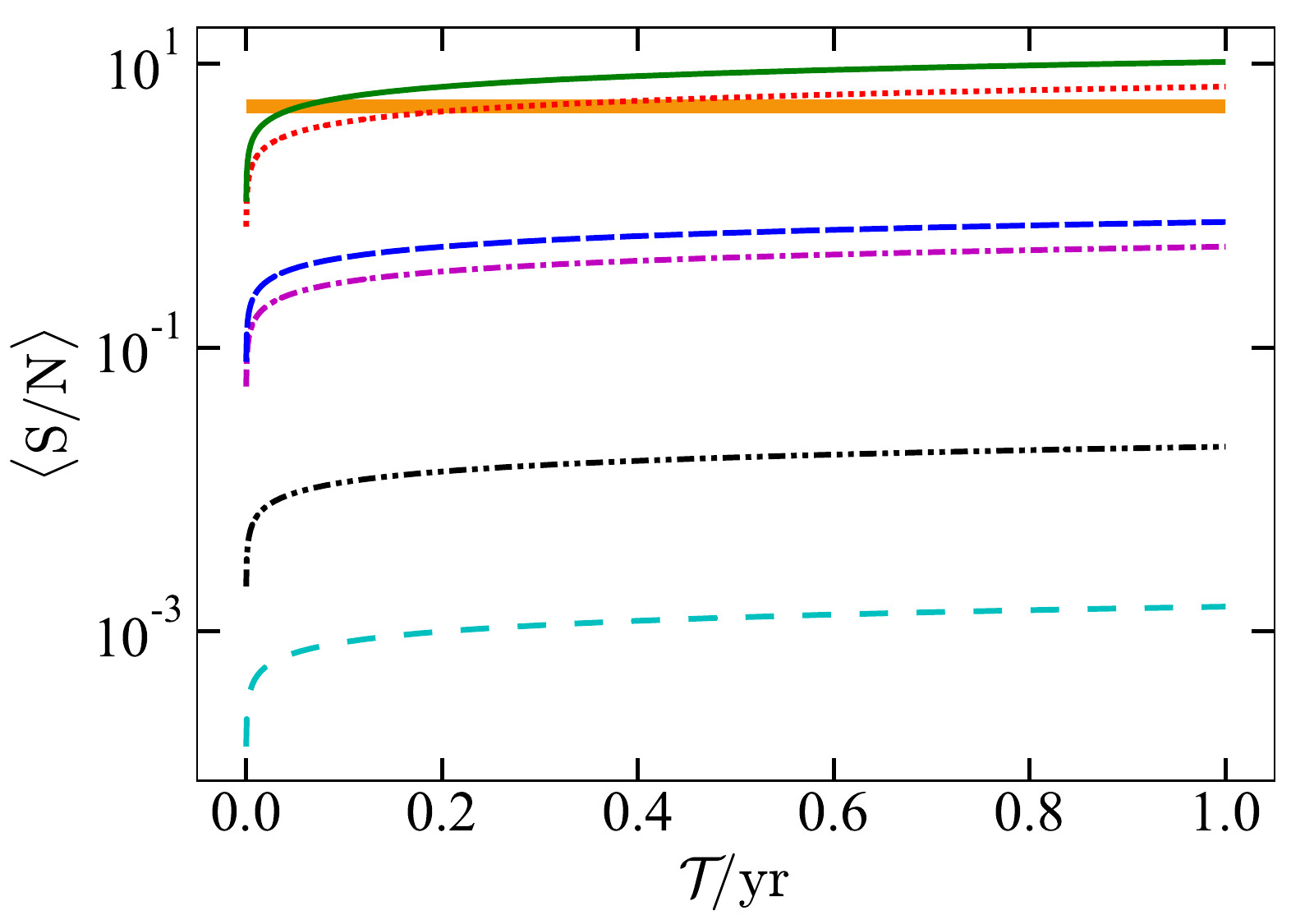}}
	\subfigure[Toroidal field dominated WD]{\includegraphics[scale=0.36]{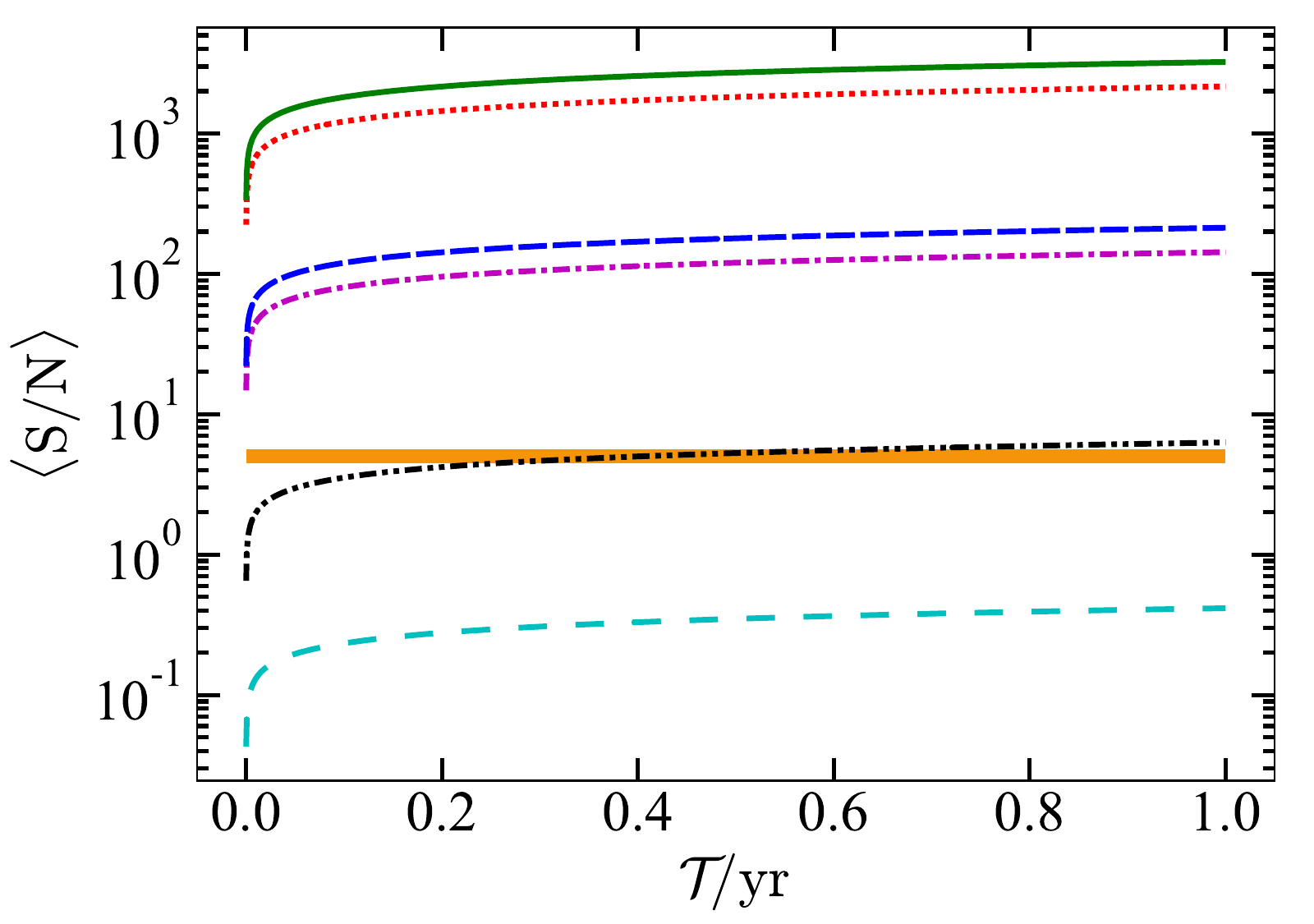}}
	\subfigure[NS with radius 14 km]{\includegraphics[scale=0.36]{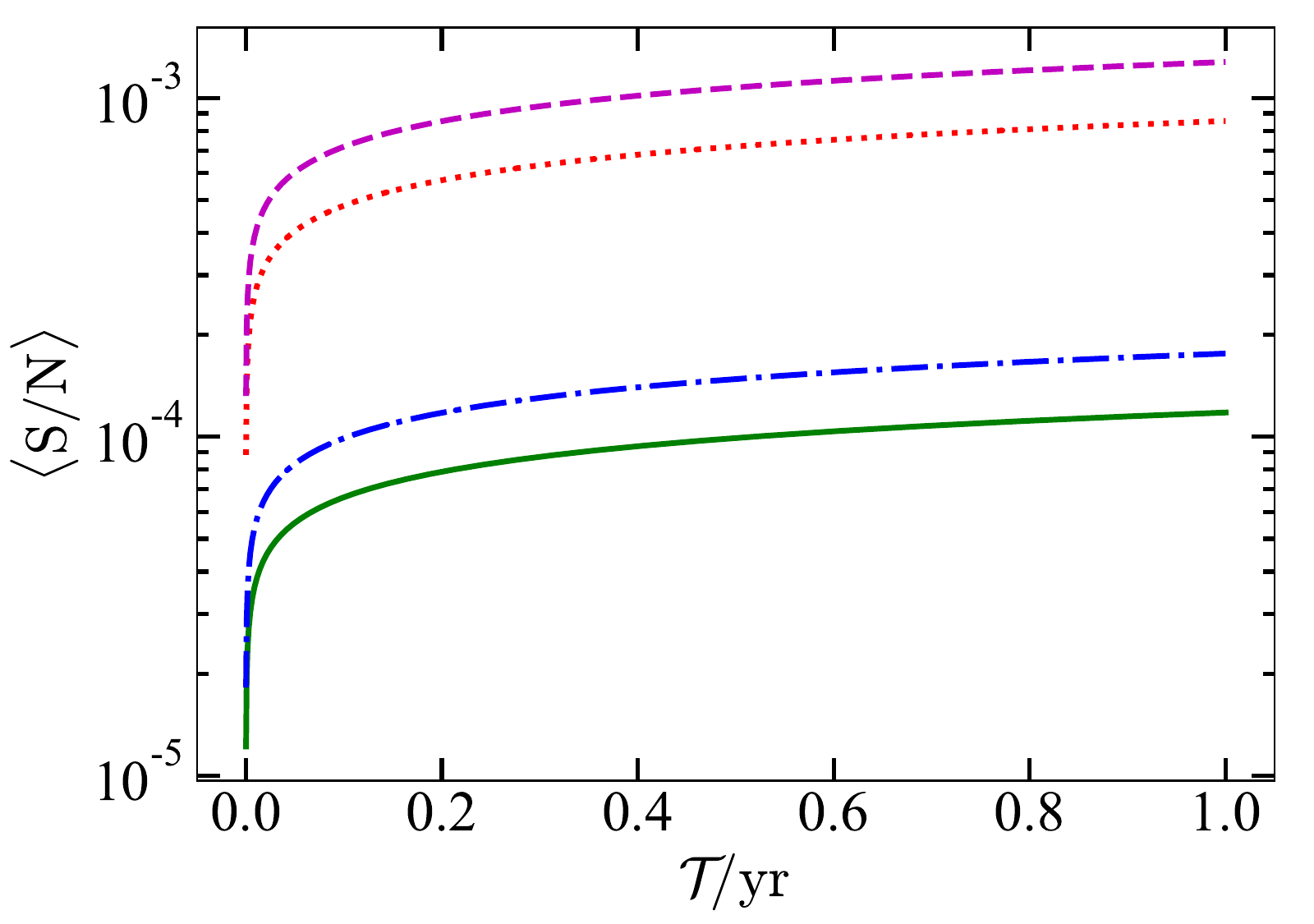}}
	\caption{As Fig.~\ref{Fig: 1RXS_J170849.0-400910} but for XTE~J$1810-197$. For parts~(a) and~(b), solid green, dotted red and double-dot-dashed black lines correspond to a WD with radius $3000\,$km rather than the $4000\,$km used previously.}
	\label{Fig: XTE_J1810-197}
\end{figure*}

We consider the SGRs and AXPs, which are neither associated with a supernova remnant nor have directly 
measured surface magnetic fields. The observed properties of these sources are listed in Table~\ref{Table: SGR 
and AXP}. The exact natures of these sources are unknown. So we separately consider them as WDs or NSs and 
obtain their structure using the {\sc xns} code  with various field configurations. Thereby we calculate their 
SNRs for various GW detectors within $1\,$yr of observation. Earlier \cite{2020MNRAS.498.4426S} considered an 
empirical formula for the ellipticity, $\epsilon = \kappa (B_\mathrm{s}^2 R_\mathrm{p}^4/G M^2)\sin^2\chi$, 
with $B_\mathrm{s}$ being the surface magnetic field and $\kappa$ a distortion parameter. However, this 
formula is valid only for poloidal magnetic field configurations. There is an effect of rotation through the 
angle $\chi$. Nevertheless, GWs are emitted if the system possesses a time-varying quadrupolar moment and the 
distortion owing to rotation cannot contribute to a time-varying quadrupolar moment 
\citep{2005ApJ...634L.165S}. So, to generate GWs, the ellipticity must contain only the effect of the magnetic 
fields and not rotation. In our calculation, we introduce the effect of magnetic fields alone to 
self-consistently calculate $\epsilon$ with the {\sc xns} code. In addition, $\kappa$ depends on the EoS of 
the star and the magnetic field configuration. However, \cite{2020MNRAS.498.4426S} chose the same $\kappa$ for 
their calculations for both WDs and NSs. We know that EoSs are very different for NSs and WDs. In principle, 
this should result in a different $\kappa$ for NSs and WDs. Using the {\sc xns} code for purely poloidal 
field, some typical parameters obtained for WDs are $\epsilon=2.93\times10^{-3}$, $M=1.41\,\rm M_\odot$, 
$R_\mathrm{p}=1220.3\,$km and $B_\mathrm{s}=8.36\times10^{11}\,$G, which leads to $\kappa \approx 9.9$. On the 
other hand, for NSs, these parameters turn out to be $\epsilon=3.6\times10^{-3}$, $M=1.58\,\rm M_\odot$, 
$R_\mathrm{p}=14.1\,$km and $B_\mathrm{s}=1.52\times10^{16}\,$G, which leads to $\kappa \approx 2.6$. Hence, 
it is evident that $\kappa$ is dependent on the EoS and it is not the same for WDs and NSs.

We now use the {\sc xns} code to obtain the structure of these SGRs and AXPs. Because we do not know the 
sources' exact nature, we run the code separately for both WD and NS EoSs with the parameters discussed in 
Section \ref{Sec: Modeling GW from a pulsar-like object}. It is to be noted that, unlike for the WD pulsars, 
the masses of these objects are unknown and so are their sizes. Hence we assume that if it is a WD its radius 
is no more than $4000\,$km and calculate two distinct classes of WD size, with polar radii $1000$ and 
$4000\,$km. Similarly, if it is a NS, we assume its circumferential polar radius to be $14\,$km. Now, from 
their $P$ and $\dot{P}$, using equation~\eqref{Eq: dipole magnetic field}, we obtain the magnetic fields at 
their poles. If the object has a dominant poloidal magnetic field, we run the code for a purely poloidal 
magnetic field configuration. On the other hand, if the object is toroidal-field dominated, we run the code 
with the maximum toroidal field inside 100~times larger than the maximum poloidal field if it is a WD or 
20~times if the source is a NS. These fields always provide stable configurations because the 
magnetic-to-gravitational energy ratios (ME/GE) are within the bounds established by 
\cite{2009MNRAS.397..763B}. Here also, for the measured $\dot{P}$, there is no significant change in 
$\Omega$ and $\chi$ within $1\,$yr. So one can
practically avoid solving equations~\eqref{Eq: radiation1} and \eqref{Eq: radiation2} for $1\,$yr of 
observation period. Obtaining the shape and size of these objects, we use equations~\eqref{Eq: SNR v21} and 
\eqref{Eq: SNR v22} to calculate the SNR for various GW detectors within $1\,$yr of integration. Moreover, 
because $B_\mathrm{p} \lesssim 10^{10}\,$G if it is a WD and $B_\mathrm{p} \lesssim 10^{14}\,$G if it is a NS, 
$L_\text{D}$ is not so large that it changes $\Omega$ significantly within $1\,$yr. Below we describe the 
time-scales for the detection of these objects one by one. Please note that \textit{LISA} and \textit{TianQin} 
cannot detect any of these sources. So we do not discuss them in this context. Moreover, any GW observations 
of SGRs and AXPs would be targeted searches just as for the WD pulsars.
\begin{itemize}
	\item 1RXS~J$170849.0-400910$: From Fig.~\ref{Fig: 1RXS_J170849.0-400910}(a), it is evident that, if this 
	is a poloidally dominated WD, no proposed detector would be able to detect it within $1\,$yr, whatever its 
	size. If it is a toroidally dominated WD, \textit{BBO} and \textit{DECIGO} could instantaneously detect it 
	as shown in Fig.~\ref{Fig: 1RXS_J170849.0-400910}(b). However, \textit{ALIA} could detect it within $4\,$d 
	if it has a radius of $4000\,$km but could not detect if its radius is $1000\,$km. From Fig.~\ref{Fig: 1RXS_J170849.0-400910}(c), it is evident that no proposed detector could detect it if it is a NS.
	\item 3XMM~J$185246.6+003317$: Fig.~\ref{Fig: 3XMM_J185246.6+003317}(a) shows that, if this is a 
	poloidally dominated WD, no proposed detector could detect it within $1\,$yr. However, if it is a 
	toroidally dominated WD with a radius of $4000\,$km, \textit{BBO} and \textit{DECIGO} would be able to 
	detect it within $1\,$d and $3\,$months, respectively (Fig.~\ref{Fig: 3XMM_J185246.6+003317}b). If it is a 
	NS, it could not be detected (Fig.~\ref{Fig: 3XMM_J185246.6+003317}c).
	\item 4U~$0142+61$: Fig.~\ref{Fig: 4U_0142+61}(a) shows the SNR as a function of time for various 
	detectors if this is a poloidal-field dominated WD when only \textit{BBO} would be able to detect it 
	within $3\,$months if it has a radius of $4000\,$km. Similarly, Fig.~\ref{Fig: 4U_0142+61}(b) shows the 
	SNR if it is a toroidal-field dominated WD. \textit{BBO} and \textit{DECIGO} could immediately detect it 
	if it has a radius $4000\,$km or within $3\,$hrs and $3\,$d, respectively, if its radius is $1000\,$km. 
	Similarly to the previous cases, if it is a NS, it could not be detected by any proposed detectors 
	(Fig.~\ref{Fig: 4U_0142+61}c).
	\item SGR~$1833-0832$: Fig.~\ref{Fig: SGR_1833-0832}(a) shows that, if this is a poloidally dominated WD 
	with a radius of $4000\,$km, \textit{BBO} and \textit{DECIGO} would be able to detect it within $3\,$d and 
	$1\,$month, respectively. If it has a dominant toroidal field (Fig.~\ref{Fig: SGR_1833-0832}b), 
	\textit{BBO} and \textit{DECIGO} could instantaneously detect it irrespective of its size, whereas 
	\textit{ALIA} would be able to detect it within $1\,$month of integration provided it has a radius 
	$4000\,$km. Note that the exact distance of this source is unknown and so we assume $d=2\,$kpc. If a 
	distance measurement is made in the future, these results can easily be manipulated because the 
	$\text{SNR} \propto 1/d$. In any case, it could not be detected if it is a NS (Fig.~\ref{Fig: SGR_1833-0832}c).
	\item XTE~J$1810-197$: We choose the maximum radius of this source to be $3000\,$km instead of $4000\,$km 
	if it is a WD because its spin is fast and the {\sc xns} code does not run for $4000\,$km with such high 
	rotation frequency. Fig.~\ref{Fig: XTE_J1810-197}(a) shows that \textit{BBO} and \textit{DECIGO} would be 
	able to detect it within $20\,$d and $100\,$d respectively, if it is a $3000\,$km poloidal-field dominated 
	WD. If it is a toroidally dominated WD, \textit{BBO} and \textit{DECIGO} could immediately detect it and 
	\textit{ALIA} would be able to detect it within $5\,$months only if it is a $3000\,$km toroidally 
	dominated WD (Fig.~\ref{Fig: XTE_J1810-197}b). On the other hand, Fig.~\ref{Fig: XTE_J1810-197}(c) shows 
	that no proposed detector would be able to detect it if it is a NS.
\end{itemize}

\subsection{SGRs and AXPs as super-Chandrasekhar WDs}

\begin{figure}
	\centering
	\includegraphics[scale=0.33]{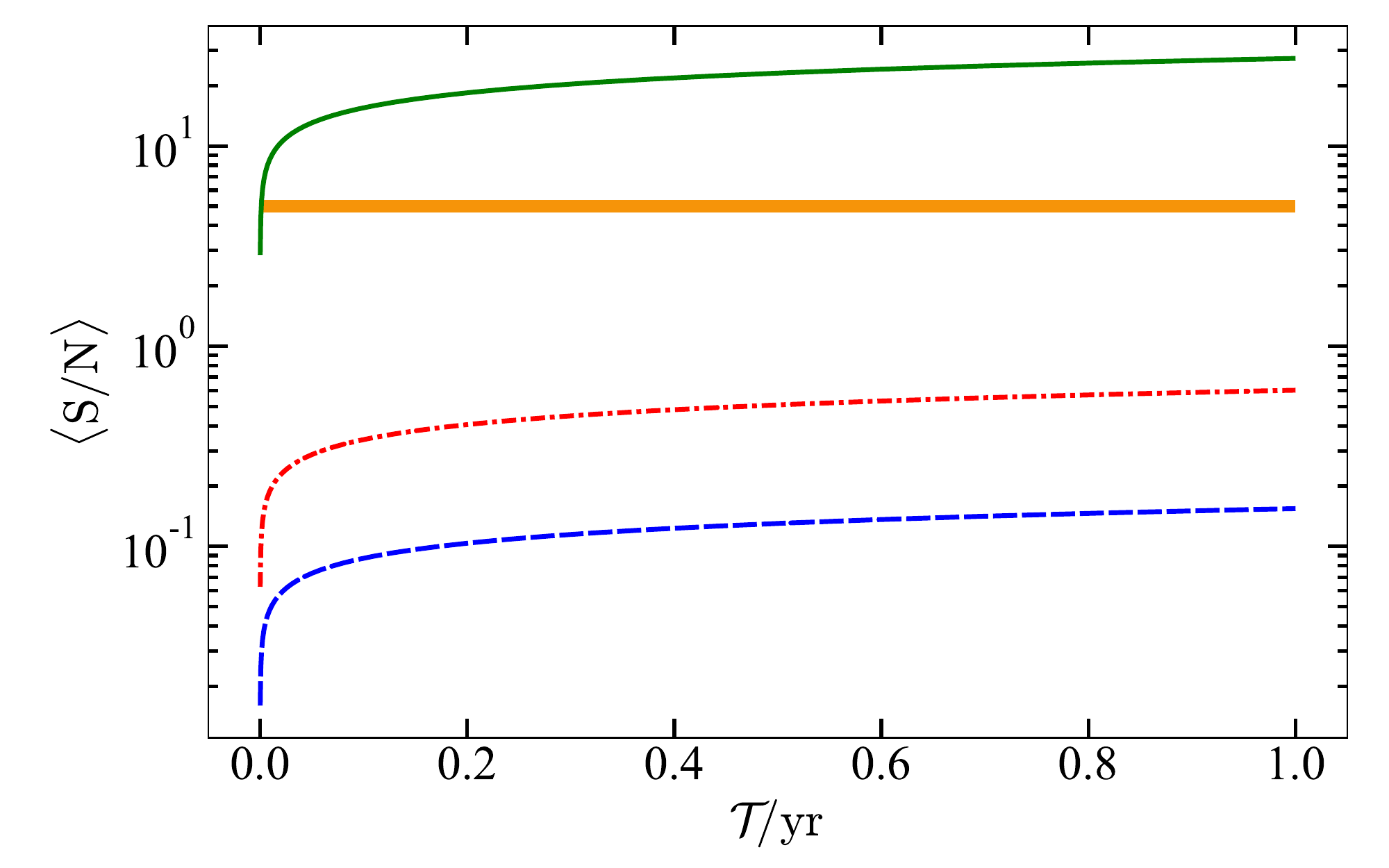}
	\caption{SNR as a function of integration time for 1RXS~J170849.0$-$400910 with $\chi=45\degree$ if it is a super-Chandrasekhar WD with $M=2\,\rm M_\odot$. Here the maximum toroidal field is $B_\mathrm{max} = 3\times 10^{14}\,$G, with magnetic-to-gravitational energy ratio of 0.19. The solid green, dot-dashed red and dashed blue lines represent respectively the \textit{ALIA}, \textit{TianQin} and \textit{LISA}. The thick orange line corresponds to $\langle\mathrm{S/N}\rangle\approx5$.}
	\label{Fig: super WD}
\end{figure}

\cite{2016JCAP...05..007M} proposed that the SGRs and AXPs can be super-Chandrasekhar WDs. Fig.~\ref{Fig: super WD} 
shows the SNR as a function of time for 1RXS~J$170849.0-400910$ with $M=2\,\rm M_\odot$. In this case, we 
consider a maximum  toroidal field inside the source of $B_\mathrm{max} = 3\times 10^{14}\,$G with a central 
density of $2\times10^{10}\,\rm g\,cm^{-3}$ and $R_\mathrm{p} = 2550\,$km. Such a configuration provides ME/GE 
$=$ 0.19, which may not give a stable equilibrium according to the criteria given by 
\cite{2009MNRAS.397..763B}. In this case, the maximum toroidal field is nearly 700 times larger than the 
maximum poloidal fields. Nevertheless, such a high field ratio is required to obtain super-Chandrasekhar WDs 
with $M\gtrsim 2\,\rm M_\odot$. \cite{2018MNRAS.477.2298Q} showed that such a ratio is indeed possible for the 
stars at the end of main sequence. From Fig.~\ref{Fig: super WD}, it is evident that such a 
super-Chandrasekhar WD could be detected by \textit{ALIA} (and hence \textit{BBO} and \textit{DECIGO}) within 
a few days of integration. However, neither \textit{LISA} nor \textit{TianQin} would be able to detect such a 
massive WD even within $5\,$yr of integration. This inference is valid for all the SGRs and AXPs in 
Table~\ref{Table: SGR and AXP} owing to $h_0$ being low for these sources because they are more distant ($d>1\,$kpc) than most WDs.

\section{Conclusions}\label{Conclusions}

SGRs, AXPs and super-Chandrasekhar WDs provide a laboratory for studying physics at high magnetic fields which 
is not yet possible to generate artificially in a terrestrial laboratory. We have considered a few general 
cases of super-Chandrasekhar WDs to explore whether future GW detectors, such as \textit{LISA}, 
\textit{TianQin}, \textit{ALIA}, \textit{BBO} and \textit{DECIGO}, would be able to detect them depending on 
their magnetic field configuration and strength. It is found that, if they have a high toroidal field, 
$L_\text{GW}$ is higher than $L_\text{D}$ and they can radiate GWs for a long time. We have determined which 
GW detectors would be able to detect these sources within $1\,$yr of observation with a large SNR. If a source 
is a poloidally dominated highly magnetized WD, only \textit{BBO} and \textit{DECIGO} would be able to detect 
it, because the rotation and magnetic axes quickly align with each other, owing to large $L_\text{D}$, and 
thereafter it no longer behaves as a pulsar. Similarly, we have estimated the detection time-scale for some of 
the known WD pulsars. From their timing properties, their surface dipole field is estimated to be 
between~$10^8$ and~$10^9\,$G, so we have found that only \textit{BBO} and \textit{DECIGO} would be able to 
detect such WD pulsars within a few months of integration time, depending on their size and magnetic field 
configuration. Of course, the emitted gravitational radiation is solely due to their spin properties and here 
we do not consider any GWs owing to the accretion from their binary partners. For some of these WD pulsars, 
such as AR~Scorpii, the exact mass is unknown and EM observations can, at best, give a mass range for them. We 
have shown here how various detectors detect these sources for various masses. In this way we can, in 
principle, estimate the exact mass of these pulsars.

We have considered those SGRs and AXPs which are not yet confirmed to be NSs. We have shown that the debate on 
the nature of these sources can easily be resolved with their CGW information. We have self-consistently 
modelled these sources using the {\sc xns} code considering all the possibilities of being toroidally or 
poloidally dominated WDs or NSs. Moreover, these sources have a spin-period $P\gtrsim 1\,$s and so, if they 
behave as pulsars, the frequency of emitted GWs falls in the range of proposed space-based GW detectors. If 
these sources are NSs, they have smaller radii and so smaller moments of inertia. As a result the GW strain is 
so small that no proposed GW detectors would be able to detect such a weak signal. However, if the sources are 
WDs, the GW strain is significantly higher and could be detected by \textit{ALIA}, \textit{BBO} and 
\textit{DECIGO} within a few hours to a few months, depending on the field configuration and strength. We have 
also shown that, if these sources are super-Chandrasekhar WDs supported by large internal magnetic fields, 
they would still be not detected by \textit{LISA} or \textit{TianQin}. \cite{2020MNRAS.498.4426S} also found a 
similar result except that, in the case of a WD, their detection time-scale is from $1$ to $5\,$yrs. This is 
because they only considered objects with poloidal fields. However, \cite{2014MNRAS.437..675W} showed that, 
inside a WD, the toroidal field can be much stronger than the poloidal field because of dynamo action at the 
time of its birth. As a result, the shape and size of the object are mostly dominated by the internal toroidal 
fields. Note that newly born NSs with large toroidal magnetic fields may be subject to secular 
instabilities \citep{2002PhRvD..66h4025C,2018MNRAS.481.4169L}. Such prolate ellipsoids experience an evolution 
where internal viscous damping of precession drives the symmetry (magnetic) axis orthogonal to the spin axis 
shortly after birth, typically within few minutes. Subsequently the external torques slowly drive $\chi$ towards $0$ 
\citep{2020MNRAS.494.4838L}. Here we address the properties of SGRs and AXPs at the current time. We try to 
address the question of whether these sources can be detected by \textit{LISA}, \textit{BBO}, \textit{DECIGO}, etc. 
when they start operating. The observed spin periods and spin-down rates indicate that the sources have been 
spinning down over hundreds of years \citep{2021ApJ...913L..12M}. Hence the internal viscous dissipation is 
insignificant at this stage.
We have shown that toroidal fields can change the shape and size of these sources more than the 
poloidal fields. So the detection time-scale for \textit{BBO} and \textit{DECIGO} decreases to a few days 
compared to a few years. This is much more productive from an observational point of view because there is an 
increase in effectiveness when the observation time-scale decreases from years to days. Overall, the detection 
of SGRs and AXPs, along with the other WD pulsars, can enhance our knowledge of compact objects' structure.

\section*{Acknowledgements}
The authors would like to thank the anonymous referee for a thorough reading of the manuscript and comments, particularly on the cumulative SNR, that have helped to improve the presentation of the work. CAT thanks Churchill College for his fellowship. TB acknowledges support from FNP through TEAM/2016-3/19 grant. BM acknowledges partial support by a project of the Department of Science and Technology (DST), India, with Grant No. DSTO/PPH/BMP/1946 (EMR/2017/001226).

\section*{Data Availability}
The data underlying this article will be shared on reasonable request to the corresponding author.


\bibliographystyle{mnras}
\bibliography{bibliography}


\bsp	
\label{lastpage}
\end{document}